\pdfoutput=1
\documentclass[doublespace, times]{stvrauth}

\usepackage{moreverb}
\usepackage[colorlinks,bookmarksopen,bookmarksnumbered,citecolor=red,urlcolor=red]{hyperref}

\newcommand\BibTeX{{\rmfamily B\kern-.05em \textsc{i\kern-.025em b}\kern-.08em
T\kern-.1667em\lower.7ex\hbox{E}\kern-.125emX}}

\usepackage{paralist}
\usepackage{subcaption}

\usepackage{todonotes}
\presetkeys{todonotes}{inline,color=yellow!40}{}

\usepackage{listings}
\usepackage{xspace}

\newcommand{\botsing}{\textrm{Botsing}\xspace}
\newcommand{\evocrash}{\textrm{EvoCrash}\xspace}
\newcommand{\evosuite}{\textrm{EvoSuite}\xspace}

\newcommand{\eg}{\textit{e.g.,~}}

\newcommand{\ie}{\textit{i.e.,~}}

\newcommand{\etal}{\textit{et al.}\xspace}

\newcommand\circled[1]{\raisebox{1.2pt}{\textcircled{\hspace{0.35pt}\scriptsize{\raisebox{-.4pt}{#1}}}}}

\usepackage{xcolor}

\def\volumeyear{2019}

\begin{document}

\runningheads{P.~Derakhshanfar, X.~Devroey, G.~Perrouin, A.~Zaidman, A.~Van Deursen}{Search-based Crash Reproduction using Behavioral Model Seeding}

\title{Search-based Crash Reproduction using Behavioral Model Seeding}

\author{Pouria Derakhshanfar\affil{1}\corrauth, 
Xavier Devroey\affil{1}, 
Gilles Perrouin\affil{2},\\
Andy Zaidman\affil{1} and
Arie van Deursen\affil{1}
}
\address{\affil{1} Delft University of Technology,
Postbus 5, 2600 AA Delft,
The Netherlands
\break
\affil{2} FNRS Research Associate, PReCISE, NADI, University of Namur, Rue de Bruxelles 61, 5000 Namur, Belgium
}

\corraddr{Delft University of Technology, Postbus 5, 2600 AA Delft, The Netherlands. E-mail: p.derakhshanfar@tudelft.nl}

\cgsn{EU Horizon 2020 ICT-10-2016-RIA ``STAMP''}{731529}

\begin{abstract}
%
Search-based crash reproduction approaches assist developers during debugging by generating a test case which reproduces a crash given its stack trace. One of the fundamental steps of this approach is creating objects needed to trigger the crash. One way to overcome this limitation is seeding: using information about the application during the search process. With seeding, the existing usages of classes can be used in the search process to produce realistic sequences of method calls which create the required objects. In this study, we introduce behavioral model seeding: a new seeding method which learns class usages from both the system under test and existing test cases.  Learned usages are then synthesized in a behavioral model (state machine). Then, this model serves to guide the evolutionary process. To assess behavioral model-seeding, we evaluate it against test-seeding (the state-of-the-art technique for seeding realistic objects) and no-seeding (without seeding any class usage). For this evaluation, we use a benchmark of 124 hard-to-reproduce crashes stemming from six open-source projects. 
Our results indicate that behavioral model-seeding outperforms both test seeding and no-seeding by a minimum of 6\% without any notable negative impact on efficiency.
\end{abstract}

\keywords{seed learning, crash reproduction, search-based software testing}

\maketitle


\section{Introduction}
\label{sec:intro}

The starting point of any debugging activity is to try to reproduce the problem reported by a user in the development environment~\cite{Zeller2009,BellerICSE2018}. In particular, for Java programs, when a crash occurs, an exception is thrown. A developer strives to reproduce it to understand its cause, then fix the bug, and finally add a (non-)regression test to avoid reintroducing the bug in future versions. 

Manual crash reproduction can be a challenging and labor-intensive task for developers: it is often an iterative process that requires setting the debugging environment in a similar enough state as the environment in which the crash occurred~\cite{Zeller2009}. Moreover, it requires the developer to have knowledge of the system's components involved in the crash. 
%
%
To help developers in their task, several automated crash reproduction methods, relying on different techniques, have been proposed \cite{Chen2015, Nayrolles2015, Nayrolles2017, Xuan2015, Bianchi2017, Soltani2017a, Soltani2018}.

One of the most promising approaches is to generate crash-reproducing test cases using search-based software testing \cite{Soltani2017a, Soltani2018}. This approach benefits from a \textit{guided genetic algorithm} which searches for the crash reproducing test case. Soltani \etal performed an empirical evaluation of \evocrash and reported that it outperforms other crash reproduction approaches~\cite{Soltani2017a}.

One of the challenges of search-based crash reproduction is to bring enough information into the test generation process.
For instance, complex elements (like strings with a particular format or objects with a complex structure) are hard to initialize without additional information. 
This can lead to two different issues: first, complex elements take more time to be generated, which can prevent to find a solution within the time budget allocated to the search; and second, elements that require complex initialization procedures (\eg specific sequences of method calls to set up an object) may prevent the search to start if the search-based approach is unable to create an initial population. 


Rojas \etal~\cite{Rojas2016} demonstrated that \emph{seeding} is beneficial for search-based unit test generation. More specifically, by analyzing source code (collecting information that relates to numeric values, string values, and class types) and existing tests (collecting information about the behavior of the objects in the test) and making them available for the search process, the overall coverage of the generated test improves. 
However, current seeding strategies focus on collecting and reusing values and object states as-is. 

In this paper, we define, implement, and evaluate a new seeding strategy, called \emph{behavioral model seeding}, which abstracts behavior observed in the source code and test cases using transition systems. The transition systems represent the (observed) usages of the classes and are used during the search to generate objects and sequences of method calls on those objects.

\emph{Behavioral model seeding} takes advantage of the advances made by the model-based testing community~\cite{Utting2007} and uses them to enhance search-based software testing. This seeding strategy helps the search process: 
\begin{inparaenum}[(i)]
\item it provides the possibility of covering the given crash by collecting information from various resources (\eg source code and existing test cases)  to infer a unique transition system; and
\item it finds the most beneficial seeding candidates, for guiding the crash reproduction search process, by defining a rational procedure for the selection of abstract object behaviors from the inferred models.
\end{inparaenum} 

%

We also adapt \emph{test seeding}, introduced by Rojas \etal, for search-based crash reproduction. Contrarily to model seeding, \emph{test seeding} relies only on the states of the objects observed during the execution of the test to seed a search process. Unlike search-based unit test generation, search-based crash reproduction does not seek to maximize the coverage of the class, but rather generates a specific test case able to reproduce a crash. Since test seeding has only been applied to search-based unit test generation~\cite{Rojas2016}, we first evaluate the use of test seeding for crash reproduction.  We then compare the results of test seeding with the application of \emph{model seeding}, which combines information on the objects states coming from the test cases with information collected in the source code, to search-based crash reproduction. 

We performed an evaluation on 124 crashes from 6 open-source applications to answer the following research questions: 
\begin{itemize}
\item[\textbf{RQ1}] What is the influence of \emph{test seeding used during initialization} on search-based crash reproduction?
\item[\textbf{RQ2}]  What is the influence of \emph{behavioral model seeding used during initialization}  on search-based crash reproduction?
\end{itemize}
We consider both research questions from the perspective of \emph{effectiveness} (of initializing the population and reproducing crashes) and \emph{efficiency}. We also investigate the factors (\eg the cost of analyzing existing tests) that influence the test and model seeding approaches and gain a better insight into how search-based crash reproduction works and how it can be improved.
Generally, our results indicate that behavioral model seeding increases the number of crashes that we can reproduce. 
More specifically, because of the randomness in the test generation process, we execute the crash replication multiple times and we observe that in the majority of these executions 4 crashes (out of 124) can be replicated; also, this seeding strategy can reproduce 9 crashes, which are not reproducible at all with no seeding, in at least one execution. In addition, this seeding strategy slightly improves the efficiency of the crash reproduction process. Moreover, model seeding enables the search process to start for three additional crashes. In contrast, using test seeding in crash-reproduction leads to a lower crash-reproduction rate and search initialization.

The contributions of this paper are:
\begin{compactenum}
    \item An evaluation of test seeding techniques applied to search-based crash reproduction;
    \item A novel behavioral model seeding approach for search-based software testing and its application to search-based crash reproduction;
    \item An open source implementation of model seeding in the \botsing toolset\footnote{Available at \url{https://github.com/STAMP-project/botsing}.}; and
    \item The discussion of our results demonstrating improvements in search-based crash reproduction abilities and contributing to a better understanding of the search-based process. All our results are available in the replication package.
\end{compactenum}

The remainder of the paper is structured as follows: Section \ref{sec:background}  provides background on search-based crash reproduction, and model-based testing. Section \ref{sec:approach} describes our behavioral model seeding strategy. Section \ref{sec:implementation} details our implementation, while Section \ref{sec:setup} explains the evaluation setup. Section \ref{sec:eval-results} presents our results. We discuss them and explain  threats to our empirical analyses in Section~\ref{sec:discussion}. Section \ref{sec:future} discusses future work and Section \ref{sec:conclusion} wraps up the paper. 

\section{Background and related work}\label{sec:background}


Application crashes that happen while the system is operating are usually reported to developer teams through an issue tracking system for debugging purposes~\cite{DBLP:conf/iwpc/WhiteVJBP15}. Depending on the amount of information reported from the operations environment, this debugging process may take more or less time. Typically, the first step for the developer is to try to reproduce the crash in his development environment~\cite{Zeller2009}. Various approaches~\cite{Bianchi2017, Chen2015, Nayrolles2017, Soltani2017a, Xuan2015} automate this process and generate a \emph{crash-reproducing test case} without requiring human intervention during the generation process.  Previous studies~\cite{Chen2015,Soltani2018a} show that such kind of test cases are helpful for the developers to debug the application.

\begin{lstlisting}[
firstnumber=0,
caption={Stack trace of the XWIKI-13372 crash},
label=lst:stacktrace,
float=t]
java.lang.NullPointerException: null
  at com[...]BaseProperty.equals([...]:96)
  at com[...]BaseStringProperty.equals([...]:57)
  at com[...]BaseCollection.equals([...]:614)
  at com[...]BaseObject.equals([...]:235)
  at com[...]XWikiDocument.equalsData([...]:4195)
  [...]
\end{lstlisting}

For Java programs, the information reported from the operations environment ideally includes a \emph{stack trace}. For instance, Listing \ref{lst:stacktrace} presents a stack trace coming from the crash XWIKI-13372.\footnote{Described in  issue \url{https://jira.xwiki.org/browse/XWIKI-13372}.} The stack trace indicates the \emph{exception} thrown (\texttt{Null\-Pointer\-Exception} here) and the \emph{frames}, \ie the stack of method calls at the time of the crash, indexed from 1 (at line 1) to 26 (not shown here).

Various approaches use a stack trace as input to automatically generate a test case reproducing the crash.
\textsc{CONCRASH}~\cite{Bianchi2017} focuses on reproducing \textit{concurrency} failures that violate thread-safety of a class by iteratively generating test code and looking for a thread interleaving that triggers a concurrency crash. 
\textsc{JCHARMING}~\cite{Nayrolles2015,Nayrolles2017} applies model checking and program slicing to generate crash reproducing tests. 
\textsc{MuCrash}~\cite{Xuan2015} exploits existing test cases written by developers. \textsc{MuCrash} selects test cases covering classes involved in the stack trace and mutate them to reproduce the crash.
%
\textsc{STAR} \cite{Chen2015} applies optimized backward symbolic execution to identify preconditions of a target crash and uses this information to generate a crash reproducing test that satisfies the computed preconditions.
Finally, \textrm{RECORE} \cite{Rossler2013} applies a search-based approach to reproduce a crash using  both a stack trace and a core dump, produced by the system when the crash happened, to guide the search.

\subsection{Search-based crash reproduction}

Search-based approaches have been widely used to solve complex, non-linear software engineering problems, which  have multiple and sometimes conflicting optimization objectives~\cite{Harman2012}. Recently, Soltani \etal~\cite{Soltani2017a} proposed a search-based approach for crash reproduction called \evocrash. \evocrash is based on the \evosuite approach \cite{Fraser2013b,Fraser2014b} and applies a new \textit{guided genetic algorithm} to generate a test case that reproduces a given crash using a distance metric, similar to the one described by Rossler \etal \cite{Rossler2013}, to guide the search.
For a given stack trace, the user specifies a \emph{target frame} relevant to his debugging activities: \ie the line with a class belonging to his system, from which the stack trace will be reproduced. For instance, applying \evocrash to the stack trace from Listing \ref{lst:stacktrace} with a target frame~2 will produce a crash-reproducing test case for the class \texttt{BaseStringProperty} that produces a stack trace with the same two first frames.

An overview of the approach is shown at the right part of Figure~\ref{fig:approach} (box 5). The first step of this algorithm, called \emph{guided initialization}, is to generate a random population. This random population is a set of random unit tests where a \emph{target method}  call (\ie the method in the target frame) is injected in each test. During the search, classical guided crossover and guided mutation are applied to the tests in such a way that they ensure that only the tests with a call to the target method are kept in the evolutionary loop.
The overall process is guided by a \emph{weighted sum fitness function} \cite{Soltani2018}, applied to each  test $t$:
\begin{equation} \label{fitness_function}
fitness(t) = 3 \times d_{l}(t) + 2 \times d_{e}(t) + d_{s}(t)
\end{equation}

The terms correspond to the following conditions when executing the test:
\begin{inparaenum}[(i)]
\item whether the execution distance from the target line ($d_{l}$) is equal to $0.0$, in which case,
\item if the target exception type is thrown ($d_{e}$), in which case,
\item if all frames, from the beginning up until the selected frame, are included in the generated trace ($d_{s}$).
\end{inparaenum}
The overall fitness value for a given test case ranges from $0.0$ (crash is fully reproduced) to $6.0$ (no test was generated), depending on the conditions it satisfies.


\subsection{Seeding strategies for search-based testing}

In addition of guided search, a promising technique is \textit{seeding}. Seeding strategies use related knowledge to help the generation process and optimize the fitness of the population~\cite{Fraser2012, Chen2018b, Lopez-Herrejon2014a}. We focus here on the usage of the source code and the available tests as primary sources of information for search-based testing. Other approaches, for instance, search for string inputs on the internet \cite{McMinn2012}, or use the existing test corpus \cite{Toffola2017} to mine relevant formatted string values (\eg XML or SQL statements).

\subsubsection{Seeding from the source code}

Three main seeding strategies are exploiting the source code for search-based testing \cite{Fraser2012, Rojas2016, Alshahwan2011}:
\begin{inparaenum}[(i)]
\item \emph{constant seeding} uses static analysis to collect and reuse constant values appearing in the source code (\eg constant values appearing in boundary conditions);
\item \emph{dynamic seeding} complements constant seeding by using dynamic analysis to collect numerical and string values, observed only during the execution of the software, and reuse them for seeding; and
\item \emph{type seeding} is used to determine the object type that should be used as an input argument, based on a static analysis of the source code (\eg by looking at \texttt{instanceof} conditions or generic types for instance).
\end{inparaenum}

\subsubsection{Seeding from the existing tests}
\label{ssec:background:testseeding}

Rojas \etal~\cite{Rojas2016} suggest two test seeding strategies, using \emph{dynamic analysis} on existing test cases: \emph{cloning} and \emph{carving}.
Dynamic analysis uses code instrumentation to trace the different methods called during an execution, which, compared to static analysis, makes it easier to identify inter-procedural sequences of method calls (for instance, in the context of a class hierarchy). Cloning and carving have been implemented in \evosuite and can be used for unit test generation.

For cloning, the execution of an existing test case is copied and used as a member of the initial population of a search process. Specifically, after its instrumentation and execution, the test case is reconstructed internally (without the assertions), based on the execution trace of the instrumented test. This internal representation is then used as-is in the initial population. Internal representation of the cloned test cases are stored in a \emph{test pool}.

For carving, an object is reused during the initialization of the population and mutation of the individuals.
In this case, only a subset of an execution trace, containing the creation of a new object and a sequence of methods called on that object, is used to internally build an object on which the methods are called. This object and the subsequent method calls are then inserted as part of a newly created test case (initialization) or in an existing test when a new object is required (mutation). Internal representations of the carved objects\footnote{In this paper, we use the term \emph{object} to refer to a carved object, \ie an object plus the sequence of methods called on that object.} are stored in an \emph{object pool}.

The integration of seeding strategies into crash reproduction is illustrated in Figure~\ref{fig:approach}, box 5.
As shown, the test cases (respectively objects) to be used by the algorithm are stored in a test case (respectively object) pool, from which they can be used according to user-defined probabilities.
For instance, if a test case only contains the creation of a new \texttt{LinkedList} (using \texttt{new}) that is  filled using two \texttt{add} method calls, the sequence, corresponding to the execution trace $<$\texttt{new}, \texttt{add}, \texttt{add}$>$, may be used as-is in the initial population (cloning) or inserted by a mutation into other test cases (carving).

\subsubsection{Challenges in seeding strategies}

The existing seeding techniques use only one resource to collect information for seeding. However, it is possible that the selected resource does not provide enough information about class usages. For instance, test seeding only uses the carved call sequences from the execution of the existing test cases. If the existing test cases do not cover the behavior of the crash in the interesting classes, this seeding strategy may even misguide the search process. Additionally, if the number of observed call sequences is large, the seeding strategy needs a procedure to prioritize the call sequences for seeding. Using random call sequences as seeds can sometimes misguide the search process. Existing seeding strategies do not currently address these issues.

\subsection{Behavioral model-based testing}

 \begin{figure}[!t]
    \centering
    \includegraphics[width=0.55\textwidth]{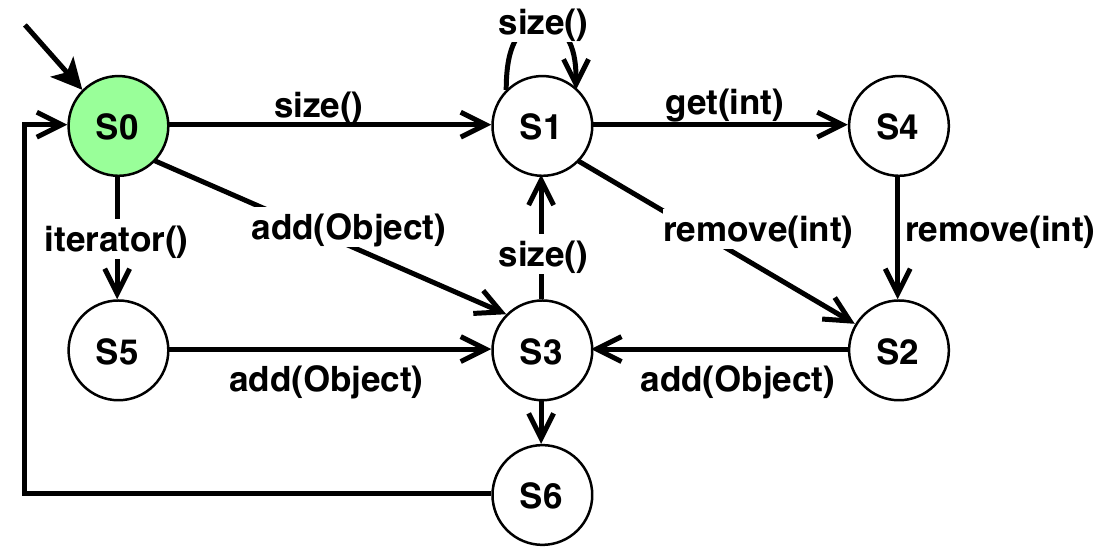}
    \caption{Transition system for method call sequences of the class \texttt{java.util.LinkedList} derived from Apache commons math source code and test cases.}
    \label{fig:list}
\end{figure}

\textit{Model-based testing} \cite{Utting2007} relies on abstract specifications (models) of the system under test to support the generation of relevant (abstract) test cases.  \textit{Transition systems} \cite{Baier2007} have been used as a fundamental formalism to reason about test case generation and support the definition of formal test selection criteria \cite{Tretmans2008}.
Each abstract test case corresponds to a sequence of method calls on one object: \ie a path in the transition system starting from the initial state and ending in the initial state, a commonly used convention to deal with finite behaviours \cite{Devroey2017b}.
Once selected from the model, abstract test cases are concretized (by mapping the transition system's paths to concrete sequences of method calls) into \emph{executable test cases} to be run on the system.
In this paper, we derive abstract test cases (called \emph{abstract object behavior} hereafter) and concretize them, producing pieces of code creating objects and invoking methods on such objects. Those pieces of code serve as seeds for search-based crash reproduction.

Figure \ref{fig:list} shows an example of a transition system representing the possible \emph{sequences} of method calls on \texttt{java.util.List} objects. Figure  \ref{fig:list} illustrates usages of methods in \texttt{java.util.List} objects, learned from the code and tests, in terms of a transition system, from which \textit{sequences} of methods calls can be derived.

The obtained transition system subsumes the behavior of the sequences used to learn it but also allows for new combinations of those sequences.  These behaviors are relevant in the context of seeding as the diversity of the objects induced is useful for the search process. Also, generating invalid behaviors from the new combinations is not a problem here as they are detectable during the search process.



\subsubsection{Abstract object behavior selection}

The abstract object behaviors are selected from the transition system according to criteria defined by the tester. In the remainder of this paper, we use \emph{dissimilarity} as selection criteria~\cite{Cartaxo2011,Hemmati2013}.
Dissimilarity selection, which aims at maximizing the fault detection rate by increasing diversity among test cases, has been shown to be an interesting and scalable alternative to other classical selection criteria \cite{Hemmati2013, Mondal2015}.
This diversity is measured using a dissimilarity distance (here, 1 - the Jaccard index \cite{Jaccard1901}) between the actions of two abstract object behaviors.

\subsubsection{Model Inference}

The model may be manually specified (and in this case will generally focus on specific aspects of the system) \cite{Utting2007}, or automatically learned from observations of the system \cite{Herbold2017, Leemans2018, Sprenkle2013, Sprenkle2011a, Tonella2014, Verwer2017}.
In the latter case, the model will be incomplete and only contain the \emph{observed behavior} of the system \cite{Tonella2012}. For instance, the sequence $<$\texttt{new}, \texttt{addAll} $>$ is valid for a \texttt{java.util.List} object but cannot be derived from the transition system in Figure \ref{fig:list} as the \texttt{addAll} method call has never been observed.
The observed behavior can be obtained via static analysis \cite{Fraser2011a} or dynamically \cite{Krka:2010:UDE:1810295.1810324}. Model inference may be used for visualization \cite{Leemans2018, Verwer2017}, system properties verification \cite{Lorenzoli2008a, Ghezzi2014}, or generation \cite{Herbold2017, Sprenkle2013, Sprenkle2011a, Prowell2004, Zhang2015a, Fraser2011a} and prioritization \cite{Devroey2017b, Dulz2003} of test cases.


 \section{Behavioral Model and Test Seeding for Crash Reproduction}
 \label{sec:approach}

The goal of behavioral model seeding (denoted model seeding hereafter) is to abstract the behavior of the software under test using models and use that abstraction during the search. At the unit test level (which is the considered test generation level in this study), each model is a transition system, like in Figure \ref{fig:list}, and represents possible usages of a class: \ie possible sequences of method calls observed for objects of that class.

\begin{figure*}[t]
    \centering
    \includegraphics[width=\textwidth]{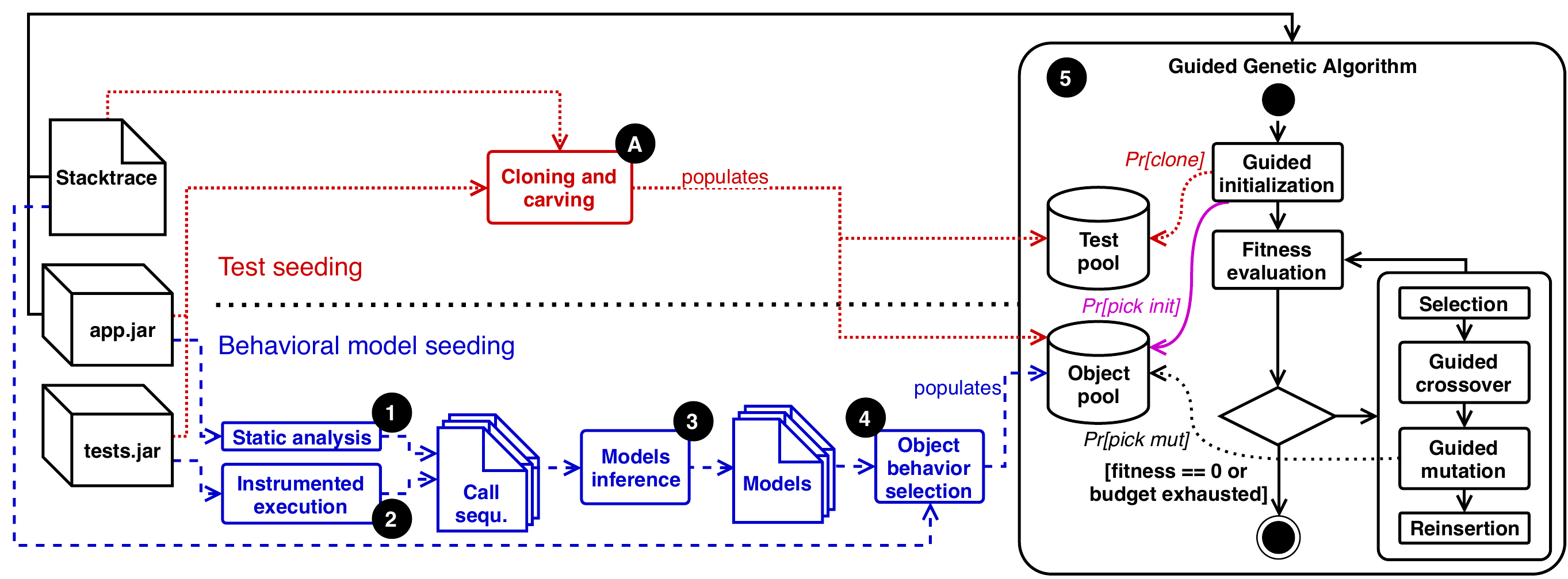}
    \caption{General overview of model seeding and test seeding for search-based crash reproduction}
    \label{fig:approach}
\end{figure*}

The main steps of our model seeding approach, presented in Figure~\ref{fig:approach}, are:
 the \emph{inference} of the individual models \circled{3} (described in Section \ref{ssec:inference}) from the \emph{call sequences} collected through \emph{static analysis} \circled{1} performed on the application code (described in Section \ref{ssec:staticanalysis}), and \emph{dynamic analysis} \circled{2} of the test cases (described in Section \ref{ssec:dynamicanalysis});
 and for each model, the \emph{selection of abstract object behaviors} \circled{4}, that are concretized into Java objects  (described in Section \ref{ssec:abstract-selection}), stored in an \emph{object pool} from which the guided genetic algorithm \circled{5}  (described in Section \ref{ssec:gga}) can randomly pick objects to build test cases during the search process.

\subsection{Model inference}
\label{ssec:inference}

Call sequences are obtained by using static analysis on the bytecode of the application \circled{1} and by instrumenting and executing the existing test cases \circled{2}.

We use $n$-gram inference to build the transition systems used for model seeding. $N$-gram inference takes a set of sequences of actions as input to produce a transition system where the $n^{th}$ action depends on the $n-1$ previously executed actions.


A large value of $n$ for the $n$-gram inference would result in wider transition systems with more states and less incoming transitions, representing a more constrained behavior and producing less diverse test cases. In contrast, a small  value of $n$ enables better diversity in the behavior allowed by the model (ending up in more diverse abstract object behaviors), requires less observations to reach stability of the model, simplifies the inference, and results in a more compact model~\cite{Sprenkle2013,Sprenkle2011a}.  For these reasons, we use $2$-gram inference to build our models. 
%
%
%

%
For each class, the model \circled{3} is obtained using a  $2$-gram inference method using the call sequences of that class.

For instance, in the transition system of Figure \ref{fig:list}, the action \texttt{size()}, executed from state $s_3$ at step $k$ only depends on the fact that the action \texttt{add(Object)} has been executed at step $k-1$, independently of the fact that there is a step $k-2$ during which the action \texttt{iterator()} has been executed.

Calls to constructors are considered as method calls during model inference. However, constructors may not appear in any transition of the model if no constructor call was observed during the collection of the call sequences.  This is usually the case when the call sequences used to infer the model have been captured from objects that are parameters or attributes of a class. If an abstract object behavior does not start by a call to a constructor, a constructor is randomly chosen to initialize the object during the concretization.


For one version of the software under test, the model inference is a one time task. Models can then be directly reused for various crash reproductions.

\subsubsection{Static analysis of the application}
\label{ssec:staticanalysis}

The static analysis is performed on the bytecode of the application. We apply this analysis to all of the available classes in the software under test.  In each method of these classes, we build the control flow graph, and for each object of that method, we collect the sequences of method calls on that object.
For each object, each path in the control flow graph will correspond to one sequence of method calls. For instance, if the code contains an \texttt{if-then-else} statement, the \texttt{true} and \texttt{false} branches will produce two call sequences. In the case of a loop statement, the  \texttt{true} branch is considered only once. The static analysis is \emph{intraprocedural}, meaning that only the calls in the current method are considered. If an object is passed as a parameter of a call to a method that (internally) calls other methods on that object, those internal calls will not appear in the call sequences.
This analysis ensures collecting all of the existing relevant call sequences for any internal or external class, which is used in the project.


\subsubsection{Dynamic analysis for the test cases}
\label{ssec:dynamicanalysis}

Since the existing manually developed test cases exemplify potential usage scenarios of the software under test, we apply dynamic analysis to collect all of the transpired sequences during the execution of these scenarios. Contrarily to static analysis, which would require an expensive effort and produce imprecise call sequences, dynamic analysis is \emph{interprocedural}. Meaning that the sequences include calls appearing in the test cases, but also internal calls triggered by the execution of the test case (\eg if the object is passed as a parameter to a method and methods are internally called on that object ). Hence, through dynamic analysis, we gain a more accurate insight into the class usages in these scenarios.

Dynamic analysis of the existing tests is done in a similar way to the carving approach of Rojas \etal~\cite{Rojas2016}: instrumentation adds log messages to indicate when a method is called, and the sequences of method calls are collected after execution. In similar fashion to static analysis, we collect call sequences of any observed object (even objects which are not defined in the software under test).
The representativeness of the collected sequences depends on the coverage of the existing tests.


\subsection{Abstract object behaviors selection}
\label{ssec:abstract-selection}

Abstract object behaviors are selected from the transition systems and concretized to populate the object pool used during the search.
To limit the number of objects in the pool, we only select abstract object behaviors from two categories of models:  models of internal classes (\ie classes belonging to packages of the software under test) and models of dependency classes (\ie classes belonging to packages of external dependencies) that are involved in the stack trace.
Since we do not seek to validate the implementation of the application, the states are ignored during the selection process.

\subsubsection{Selection}
\label{ssec:selection}

There exist various criteria to select abstract object behaviors from transition systems~\cite{Utting2007}.
To successfully guide the search, we need to establish a good ratio between \emph{exploration} (the ability to visit new regions of the search space) and \emph{exploitation} (the ability to visit the neighborhood of previously visited regions) \cite{vcrepinvsek2013}.
The guided genetic operators which are introduced in \evocrash approach \cite{Soltani2017a} guarantee the exploitation by focusing the search based on the methods in the stack trace. However, depending on the stack trace, focusing on particular methods may reduce the exploration. Poor exploration decreases the diversity of the generated tests and may trap the search process in local optima.

To improve the exploration ability in the search process, we use \emph{dissimilarity} as the criterion to select the abstract object behaviors. Compared to classical structural coverage criteria that seek to cover as many parts of the transition system as possible, dissimilarity tries to increase diversity among the test cases by maximizing a distance $d$ (\ie the Jaccard index \cite{Jaccard1901}):
$$
d = 1 - \frac{ \{call_{1i} \in b_{1}\} \cap \{call_{2j} \in b_{2}\}}{\{call_{1i} \in b_{1}\} \cup \{call_{2j} \in b_{2}\}}
$$
Where $b_{1} = <\mathit{call}_{11}, \mathit{call}_{12}, \ldots> $ and $ b_{2} = <\mathit{call}_{21}, \mathit{call}_{22}, \ldots>$ are two abstract object behaviors.

\subsubsection{Concretization}
\label{ssec:concretization}

Each abstract object behavior has to be concretized to an object and method calls before being added to the objects pool. In other words, for each abstract object behavior, if the constructor invocation is not the first action, one constructor is randomly called; and the methods are called on this object in the order specified by the abstract object behavior with randomly generated parameter values. Due to the randomness, each concretization may be different from the previous one. For each abstract object behavior, $n$ concretizations (default value is $n=1$  to balance scalability and diversity of the objects in the object pool) are done for each abstract object behavior and saved in the object pool.
For instance, Listing \ref{lst:concretized-test} shows the concretized abstract object behavior {\ttfamily\footnotesize <add(Object), add(Object)>} derived from the transition system model of Figure \ref{fig:list}. The type of the parameters (\texttt{EuclideanIntegerPoint}) is randomly selected during the concretization and created with required parameter values (an integer array here).

\begin{lstlisting}[
language=Java,
caption={Concretized abstract object behavior for \texttt{LinkedList} based on the transition system model of Figure \ref{fig:list}},
label=lst:concretized-test,
float=t]
int[] t = new int[7];
t[3] = -2147483647;
EuclideanIntegerPoint ep = new EuclideanIntegerPoint(t);
LinkedList<[...]> lst = new LinkedList<>();
lst.add(ep);
lst.add(ep);
\end{lstlisting}

\subsection{Guided Initialization and Guided Mutation}
\label{ssec:gga}

Classes are instantiated to create objects during two main steps of the guided genetic algorithm: guided initialization, where objects are needed to create the initial set of test cases; and guided mutation, where objects may be required as parameters when adding a method call. When no seeding is used, those objects are randomly created (as in the concretization step described in Section~\ref{ssec:concretization}) by calling the constructor and random methods.

Finally, to preserve exploration in model seeding, objects are picked from the object pool during guided initialization (resp. guided mutation) according to a user-defined probability $Pr[pick\ init]$ (resp. $Pr[pick\ mut]$), and randomly generated otherwise.
In our evaluation, we considered four different values for $Pr[pick\ init] \in \{0.2, 0.5, 0.8, 1.0\}$, to study the effect of model seeding on the initialization of the search process. Furthermore, we fixed the value of $Pr[pick\ mut] = 0.3$, corresponding to the default value of $Pr[pick\ mut]$ for test seeding for classical unit test generation in \evosuite.

\begin{lstlisting}[
    numbers=left,
    caption={Stack trace excerpt for MATH-79b},
    label=lst:MATH-79b-trace,
    float=t]
    java.lang.NullPointerException
     at ...KMeansPlusPlusClusterer.assignPointsToClusters()
     at ...KMeansPlusPlusClusterer.cluster()
    \end{lstlisting}

\begin{lstlisting}[
language=Java,
escapechar=|,
caption={Test generated for frame 2 of MATH-79b (Listing \ref{lst:MATH-79b-trace})},
label=lst:generated-test,
float=t]
public void testCluster() throws Exception{
 int[] t = new int[7]; | \label{line:startobject} |
 t[3] = (-2147483647);
 EuclideanIntegerPoint ep = new EuclideanIntegerPoint(t);
 LinkedList<[...]> lst = new LinkedList<>();
 lst.add(ep);
 lst.add(ep); | \label{line:endobject} |
 KMeansPlusPlusClusterer<[...]> kmean = new KMeansPlusPlusClusterer<>(12);
 lst.offerFirst(ep); | \label{line:additional} |
 kmean.cluster(lst, 1, (-1357));} | \label{line:target} |
\end{lstlisting}

As an example of object picking in action, test case generation with model seeding generated the test case in Listing \ref{lst:generated-test} for the second frame of the stack trace from the crash MATH-79b from the Apache commons math project, reported in Listing \ref{lst:MATH-79b-trace}.
The target method is the last method called in the test (line \ref{line:target}) and throws a \texttt{NullPointerException}, reproducing the input stack trace. The first parameter of the method has to be a \texttt{Collection<T>} object. In this case, the guided genetic algorithm picked the list object from the object pool (from Listing \ref{lst:concretized-test}) and inserted it in the test case (lines \ref{line:startobject} to \ref{line:endobject}). The algorithm also modified that object (during guided mutation) by invoking an additional method on the object (line \ref{line:additional}).

 \subsection{Test seeding}

As described in Section \ref{ssec:background:testseeding}, test seeding starts by executing the test cases (Figure \ref{fig:approach} box \circled{A}) for carving and cloning, and subsequently populating the test and object pools. Like for model seeding, only internal classes and external classes appearing in the stack trace are considered.

For crash reproduction, the test pool is used only during guided initialization to clone test cases that contain the target class, according to a user-defined $Pr[clone]$ probability. If the target method is not called in the cloned test case, the guided initialization also mutates the test case to add a call to the target method.
The object pool is used during the guided initialization and guided mutation to pick objects. 
As described by Rojas \etal~\cite{Rojas2016}, the properties of using the object pool during initialization ($Pr[pick\ init]$) and mutation ($Pr[pick\ mut]$) are indicated as a single property called \texttt{p\_object\_pool} in test seeding.

\section{Implementation}\label{sec:implementation}



Relying on the \evocrash experience \cite{Soltani2017a, Soltani2018a, Soltani2019},  we developed \botsing, a framework for crash reproduction with extensibility in mind.
 \botsing also relies on \evosuite~\cite{Fraser2011} for the code instrumentation during test generation and execution by using \textit{evosuite-client} as a dependency.
Our open-source implementation is available at \url{https://github.com/STAMP-project/botsing}. The current version of \botsing includes both test seeding and model seeding as features.

\subsection{Test seeding}

Test seeding relies on the implementation defined by Rojas \etal~\cite{Rojas2016} and available in \evosuite. This implementation requires the user to provide a list of test cases to consider for cloning and carving. In \botsing, we automated this process using the dynamic analysis of the test cases to automatically detect those accessing classes involved in a given stack trace. We also modified the standard guided initialization and guided mutation to preserve the call to the target method during cloning and carving.

\subsection{Model seeding}

As mentioned in Section~\ref{sec:approach}, \botsing uses a combination of static and dynamic analysis to infer models. The static analysis (\circled{1} in Figure \ref{fig:approach}) uses the reflection mechanisms of \evosuite to inspect the compiled code of the classes involved in the stack traces, and collect call sequences. The dynamic analysis (\circled{2} in Figure \ref{fig:approach}) relies on the test seeding mechanism used for cloning that allows inspecting an internal representation of the test cases obtained after their execution and collect call sequences.
The resulting call sequences are then used to infer the transition system models of the classes using a 2-gram inference tool called YAMI~\cite{Devroey2017b} (\circled{3} in Figure \ref{fig:approach}).
From the infered models, we extract a set of dissimilar (based on the Jaccard distance \cite{Jaccard1901}) abstract object behaviors (\circled{4} in Figure \ref{fig:approach}). For abstract object behavior extraction, we use the VIBeS~\cite{Devroey2016} model-based testing tool.
Abstract object behaviors are then concretized into real objects. For this concretization, we rely on the EvoSuite API.

\section{Empirical Evaluation}\label{sec:setup}

Our evaluation aims to assess the effectiveness of each of the mentioned seeding strategies (model and test seeding) on search-based crash reproduction.
For this purpose, first, we evaluate the impact of each seeding strategy on the number of reproduced crashes. Second, we examine if using each of these strategies leads to a faster crash reproduction. Third, we see if each seeding strategy can help the search process to start more often. Finally, we characterize the impacting factors of test and model seeding.

Since the focus of this study is using seeding to enhance the guidance of the search initialization, we examine different probabilities of using the seeded information during the guided initialization in the evaluation of each strategy. Hence, we repeat each execution of test seeding with the following values for $Pr[clone]$: 0.2, 0.5, 0.8, and 1.0. Likewise, we execute each execution of model seeding with the same values for $Pr[pick\ init]$ (which is the only property that we can use for modifying the probability of the object seeding in the initialization of model seeding).


\subsection{Research questions}
\label{sec:setup:rqs}

In order to assess the usage of test seeding applied to crash reproduction and our new model seeding approach during the guided initialization, we performed an empirical evaluation to answer the two research questions defined in Section \ref{sec:intro}.

\textbf{RQ1} \emph{What is the influence of test seeding used during initialization on search-based crash reproduction?}
To answer this research question, we compare \botsing executions with \emph{test seeding} enabled to executions where no additional seeding strategy is used (denoted \emph{no seeding} hereafter), from their effectiveness to reproduce crashes and start the search process, the factors influencing this effectiveness, and the impact of test seeding on the efficiency. We divide \textbf{RQ1}  into four sub-research questions:
\begin{compactitem}
    \item[\textbf{RQ1.1}] Does test seeding help to reproduce more crashes?
    \item[\textbf{RQ1.2}] Does test seeding impact the efficiency of the search process?
    \item[\textbf{RQ1.3}] Can test seeding help to initialize the search process? 
    \item[\textbf{RQ1.4}] Which factors in test seeding impact the search process?
\end{compactitem}

\textbf{RQ2}  \emph{What is the influence of behavioral model seeding used during initialization on search-based crash reproduction?}
To answer this question, we compare \botsing executions with \emph{model seeding} to executions with \emph{test seeding} and \emph{no seeding}. We also divide \textbf{RQ2}  into four sub-research questions:
\begin{compactitem}
    \item[\textbf{RQ2.1}] Does behavioral model seeding help to reproduce more crashes compared to no seeding?

    \item[\textbf{RQ2.2}] Does behavioral model seeding impact the efficiency of the search process compared to no seeding?
    \item[\textbf{RQ2.3}] Can behavioral model seeding help to initialize the search process compared to no seeding? 
    \item[\textbf{RQ2.4}] Which factors in behavioral model seeding impact the search process?
\end{compactitem}

\subsection{Setup}
\label{sec:setup:setup}

\subsubsection{Crash selection}
\label{sec:setup:setup:selection}
In a recent study about the evaluation of search-based crash reproduction approaches, Soltani \etal \cite{Soltani2019} introduced a new benchmark, called JCrashPack, containing 200 real-world crashes from seven projects: \textit{JFreeChart}, \textit{Commons-lang}, \textit{Commons-math}, \textit{Mockito}, \textit{Joda-time}, \textit{XWiki}, and \textit{ElasticSearch}.
 We use the same benchmark for the empirical evaluation of model-seeding and test-seeding on crash reproduction.

To use test and model seeding for reproducing the crashes of JCrashPach, first, we needed to apply static and dynamic analysis on different versions of projects in this benchmark. We successfully managed to run static analysis on all of the classes of JCrashPack.
On the contrary, we observed that dynamic analysis was not successful in the execution of existing test suites of ElasticSearch. The reason for this failure stemmed from the technical difficulty of running ElasticSearch tests by the EvoSuite test executor. 
Since both of the seeding strategies need dynamic analysis, we excluded ElasticSearch cases from JCrashPack for this experiment. 
JCrashPack contains 124 crashes after excluding ElasticSearch cases. Table~\ref{tab:projects} provides more details about our dataset.

We used the selected crashes for the evaluation of \emph{no seeding} and \emph{model seeding}. Since \emph{test seeding} needs existing test cases that are using the target class, we filtered out the crashes which contain only classes without any using tests. Hence, we used only 59 crashes for the evaluation of \emph{test seeding}. More information about average number of used test classes for test seeding is available in Table \ref{tab:seeding-info}.

\begin{table} [t]
    \centering
	\caption{Projects used for the evaluation with the number of crashes (\textbf{Cr.}), the average number of frames per stack trace ($\overline{\mathbf{frm}}$), the average cyclomatic complexity ($\overline{\mathbf{CCN}}$), the average number of statements ($\overline{\mathbf{NCSS}}$), the average line coverage of the existing test cases ($\overline{\mathbf{LC}}$), and the average branch coverage of the existing test cases ($\overline{\mathbf{BC}}$).}
    \begin{footnotesize}
	\begin{tabular}{l rrrrrr}
\textbf{Application} & \textbf{Cr.} & $\overline{\mathbf{frm}}$ & $\overline{\mathbf{CCN}}$ & $\overline{\mathbf{NCSS}}$& $\overline{\mathbf{LC}}$ & $\overline{\mathbf{BC}}$ \\ 
\hline 
JFreeChart            &   2 & 6.00 & 2.75 & 63.01k & 67\% & 59\% \\ 
Commons-lang    &   22 & 2.04 & 3.28 & 13.38k & 91\% & 87\% \\ 
Commons-math   &   27 & 3.92 & 2.43 & 29.98k & 90\% & 84\% \\ 
Mockito                 &   14 & 4.64 & 1.78 & 6.06k & 97\% & 93\% \\ 
Joda-Time            &   8 & 3.87 & 2.11 & 19.41k & 89\% & 82\% \\ 
XWiki                    &  52 & 27.25 & 1.92 & 177.84k & 73\% & 71\% \\ 
\end{tabular}
	\end{footnotesize}
	\label{tab:projects}
\end{table}

\begin{table}[t]
    \center
    \caption{Information about test classes and models used, respectively, for test and model seeding in each project. $\overline{test}$ designate the average number of test classes used for test seeding. Also, $\overline{state}$, $\overline{trans}$, and $\overline{BFS}$ denote the average number of states, transitions, and BFS height of the used models, respectively. The standard deviations of each of these metrics ($\sigma$) are located beside them.}
    \label{tab:seeding-info}
    \tiny
    \begin{tabular}{ l | r r}
\hline 
\textbf{Project} & \textbf{$\overline{test}$} & \textbf{$\sigma$} \\ 
\hline 
chart  &  29.17  &  20.01\\ 
lang  &  1.45  &  2.03\\ 
math  &  1.24  &  1.37\\ 
mockito  &  0.73  &  2.15\\ 
time  &  9.24  &  9.55\\ 
xwiki  &  0.14  &  1.09\\ 
\end{tabular}
    \begin{tabular}{ l | r r r r r r}
\hline 
\textbf{Project} & \textbf{$\overline{state}$} & \textbf{$\sigma$} & \textbf{$\overline{trans}$} & \textbf{$\sigma$} & \textbf{$\overline{BFS}$} & \textbf{$\sigma$} \\ 
\hline 
chart  &  56.67  &  50.40  &  157.50  &  167.86  &  21.00  &  17.50\\ 
lang  &  39.69  &  51.49  &  117.96  &  158.07  &  5.58  &  7.32\\ 
math  &  14.00  &  12.46  &  34.22  &  40.59  &  5.20  &  4.11\\ 
mockito  &  12.18  &  10.93  &  21.45  &  22.70  &  5.32  &  3.90\\ 
time  &  63.35  &  40.85  &  230.80  &  167.99  &  16.10  &  11.79\\ 
xwiki  &  47.94  &  90.94  &  139.15  &  323.75  &  11.08  &  17.04\\ 
\end{tabular}
\end{table}

\subsubsection{Model inference}
Since the selected crashes for this evaluation are identified before the model inference process, we have applied the dynamic analysis only on the test cases which use the involved classes in the crashes. During the static analysis, we spot all relevant test cases which call the methods of the classes that have appeared in the stack traces of the crashes. Next, we apply dynamic analysis only on the detected relevant test cases.
This filtering process helps us to shorten the model inference execution time without losing accuracy in the generated models.

More information about the inferred models is available in Table \ref{tab:seeding-info}.

\subsubsection{Configuration parameters}

We used a budget of 62,328 fitness evaluations (corresponding on average to 15 minutes of executing \botsing with no seeding on our infrastructure which is introduced in section \ref{sec:setup:setup:infrst}) to avoid side effects on execution time when executing \botsing on different frames in parallel.
We also fixed the population size to 100 individuals as suggested by the latest study on search-based crash reproduction~\cite{Soltani2018}.
All other configuration parameters are set at their default value~\cite{Rojas2016}, and we used the default weighted sum scalarization fitness function (Equation \ref{fitness_function}) from Soltani \etal \cite{Soltani2018}.

For test seeding executions, as we described at the beginning of this section, we execute each execution with four values for $Pr[clone]$: 0.2 (which is the default value), 0.5, 0.8, and 1.0. Also, we used the default value of 0.3 for \texttt{p\_object\_pool}.



We also use values  0.2, 0.5, 0.8, and 1.0 for $Pr[pick\ init]$ for model seeding executions. The value of $Pr[pick\ mut]$, which indicates the probability of using seeded information during the mutation, is fixed at 0.3. In addition to model seeding configurations, we fix the size of the selected abstract object behaviors to the size of the individual population in order to ensure that there are enough test cases to initiate the search.



For each frame (951 in total), we executed \botsing for \emph{no seeding} (i.e., no additional seeding compared to the default parameters of \botsing) and each configuration of \emph{model seeding}. 
Since \emph{test seeding} needs existing test cases which are using the target class, we filtered out the frames that do not have any test for execution of this seeding strategy. Therefore, we executed each configuration of \emph{test seeding} on the subset of frames (171 in total).

\subsubsection{Infrastructure}
\label{sec:setup:setup:infrst}
We used 2 clusters (with 20 CPU-cores, 384 GB memory, and 482 GB hard drive) for our evaluation. For each stack trace, we executed an instance of \botsing for each frame which points to a class of the application. We discarded other frames to avoid generating test cases for external dependencies. 
We ran \botsing on 951 frames from 124 stack traces for no-seeding and each model-seeding strategy configuration. Also, we ran \botsing with test-seeding on 171 frames from 59 crashes. To address the random nature of the evaluated search approaches, we repeated each execution 30 times.
We executed a total of 186,560 independent executions for this study. These executions took about 18 days overall.

\subsection{Data analysis procedure} 
\label{sec:setup:analyzing}

To check if the search process can reach a better state using seeding strategies, we analyze the status of the search process after executing each of the cases (each run in one frame of a stack trace). We define 5 states: 
\begin{compactenum}[(i)]
\item \textbf{not started}, the initial population could not be initialized, and the search did not start;
\item \textbf{line not reached}, the target line could not be reached;
\item \textbf{line reached}, the target line has been reached, but the target exception could not be thrown;
\item \textbf{ex. thrown}, the target line has been reached, and an exception has been thrown but produced a different stack trace; and 
\item \textbf{reproduced} the stack trace could be reproduced.
\end{compactenum}
Since we repeat each execution 30 times, we use the majority of outcomes for a frame reproduction result. For instance, if \botsing reproduces a frame in the majority of the 30 runs, we count that frame as a \textit{reproduced}.

To measure the impact of each strategy in the crash reproduction ratio (\textbf{RQ1.1} and \textbf{RQ2.1}), we use the Odds Ratio (OR) because of the binary distribution of the related data:  a search process either reproduces a crash (the generated test replicates the stack trace from the highest frame which is reproduced by at least one of the other searches) or not. Also, we apply Fisher's exact test, with $\alpha = 0.05$ for the Type I error, to evaluate the significance of results.

Moreover, to answer \textbf{RQ1.2} and \textbf{RQ2.2}, which investigate the efficiency of the different strategies, we compare the number of fitness function evaluations needed by the search to reach crash reproduction.
This metric indicates if seeding strategies lead to better initial populations that need fewer iterations to achieve the crash reproducing test. Since efficiency is only relevant for the reproduced cases, we only applied this comparison on the crashes which are reproduced at least once by no seeding or the seeding strategy (test seeding for RQ1.2 and model seeding for RQ2.2).
 We use the Vargha-Delaney statistic \cite{vargha} to appraise the effect size between strategies. In this statistic, a value lower than 0.5 for a pair of factors $(A,B)$ gives that $A$ reduces the number of needed fitness function evaluations, and a value higher than 0.5 shows the opposite. Also, we use the Vargha Delaney magnitude measure to partition the results into three categories having large, medium, and small impact. In addition, to examine the significance of the calculated effect sizes, we use the non-parametric Wilcoxon Rank Sum test, with  $\alpha = 0.05$  for Type I error. Moreover, we do note that since the reproduction ratio of each strategy is not 30/30 for each crash, executions that could not reproduce the frame simply reached the maximum allowed budget (62,328).

To measure the impact of each strategy in initializing the first population (\textbf{RQ1.3} and \textbf{RQ2.3}), we use the same procedure as \textbf{RQ1.1} and \textbf{RQ2.1} because the distribution of related data in this aspect is binary too (i.e., whether the search process can start the search or not).

For all of the statistical tests in this study, we only use a level of significance $\alpha = 0.05$.

Since the model inference (in model seeding) and test carving (in test seeding) techniques can be applied as one time processes before running any search-based crash reproduction, we do not include them in the efficiency evaluation.

To answer \textbf{RQ1.4} and \textbf{RQ2.4}, we performed a manual analysis on the logs and crash reproducing test case (if any). We focused our manual analysis on the crash reproduction executions for which the search in one seeding configuration has a significant  impact (according to the results of the previous sub-research questions) on 
\begin{inparaenum}[(i)]
\item \textit{initializing the initial population},
\item \textit{crash reproduction},
\item or \textit{search process efficiency}
\end{inparaenum}
 compared to no-seeding.
Based on our manual analysis, we used a card sorting strategy by assigning keywords to each frame result and grouping those keywords to identify influencing factors.

\section{Evaluation Results}
\label{sec:eval-results}

We present the results of the evaluation and answer the two research questions by comparing each seeding strategy with no-seeding.
\subsection{Test seeding (RQ1)}


\begin{table}[t]
    \center
    \caption{Odds ratios of model/test seeding configurations vs. no seeding in crash reproduction ratio. This table only shows the crashes, which reveal statistically significant differences (p-value $< 0.05$). An Odds ratio value higher than 1.0 gives that the seeding strategy is better than no seeding, and a value lower than 1.0 shows the opposite.}
	\label{tab:oddratios}
    \tiny
    \begin{tabular}{ l | l | r}
\hline 
\textbf{Conf.} & \textbf{Crash} & \textbf{Odds Ratio (p-value)} \\ 
\hline 
test s. 0.5 & LANG-6b & Inf (2.37e-02)\\ 
 & MATH-1b & 0.00 (1.69e-17)\\ 
 & MATH-61b & 0.00 (1.69e-17)\\ 
 & CHART-4b & 0.00 (1.69e-17)\\ 
 & TIME-20b & 0.00 (1.94e-03)\\ 
 & TIME-10b & 207.79 (2.36e-12)\\ 
 & TIME-5b & 3.52 (3.52e-02)\\ 
\hline 
test s. 0.8 & LANG-6b & Inf (1.94e-03)\\ 
 & MATH-1b & 0.00 (1.69e-17)\\ 
 & MATH-61b & 0.00 (1.69e-17)\\ 
 & CHART-4b & 0.00 (1.69e-17)\\ 
 & TIME-20b & 0.00 (4.64e-05)\\ 
 & TIME-10b & Inf (9.23e-14)\\ 
 & TIME-7b & 0.00 (6.19e-07)\\ 
\hline 
test s. 1.0 & LANG-51b & 0.21 (8.21e-03)\\ 
 & LANG-6b & Inf (4.64e-05)\\ 
 & MATH-1b & 0.00 (1.69e-17)\\ 
 & MATH-61b & 0.00 (1.69e-17)\\ 
 & CHART-4b & 0.00 (1.69e-17)\\ 
 & TIME-20b & 0.00 (5.83e-06)\\ 
 & TIME-10b & 69.79 (2.82e-10)\\ 
\hline 
test s. 0.2 & MATH-1b & 0.00 (1.69e-17)\\ 
 & MATH-61b & 0.00 (1.69e-17)\\ 
 & CHART-4b & 0.00 (1.69e-17)\\ 
 & TIME-20b & 0.00 (3.19e-04)\\ 
 & TIME-10b & Inf (9.23e-14)\\ 
 & TIME-7b & 0.00 (1.05e-02)\\ 
\hline 
\end{tabular}
    \begin{tabular}{ l | l | r}
\hline 
\textbf{Conf.} & \textbf{Crash} & \textbf{Odds Ratio (p-value)} \\ 
\hline 
model s. 0.2 & LANG-9b & Inf (1.94e-03)\\ 
 & LANG-51b & 0.17 (3.33e-03)\\ 
 & MOCKITO-10b & Inf (1.43e-08)\\ 
 & XWIKI-13141 & 13.95 (5.58e-03)\\ 
\hline 
model s. 0.5 & LANG-9b & Inf (2.37e-02)\\ 
 & MOCKITO-10b & Inf (1.87e-07)\\ 
 & XWIKI-13141 & Inf (7.97e-04)\\ 
 & XWIKI-14152 & 6.66 (7.41e-03)\\ 
\hline 
model s. 0.8 & LANG-9b & Inf (1.94e-03)\\ 
 & LANG-51b & 0.29 (3.70e-02)\\ 
 & MOCKITO-10b & Inf (8.27e-10)\\ 
 & XWIKI-13141 & Inf (7.97e-04)\\ 
 & XWIKI-14152 & 11.24 (2.51e-04)\\ 
\hline 
model s. 1.0 & LANG-9b & Inf (1.94e-03)\\ 
 & MOCKITO-10b & Inf (5.34e-08)\\ 
 & XWIKI-13141 & 13.95 (5.58e-03)\\ 
 & XWIKI-14152 & 32.80 (5.62e-08)\\ 
\hline 
\end{tabular}
\end{table}

\subsubsection{Crash reproduction effectiveness (\textbf{RQ1.1})}

\begin{figure*}[t]
	\centering
	\begin{subfigure}[t]{0.48\textwidth}
	    \centering
	    \includegraphics[height=38mm]{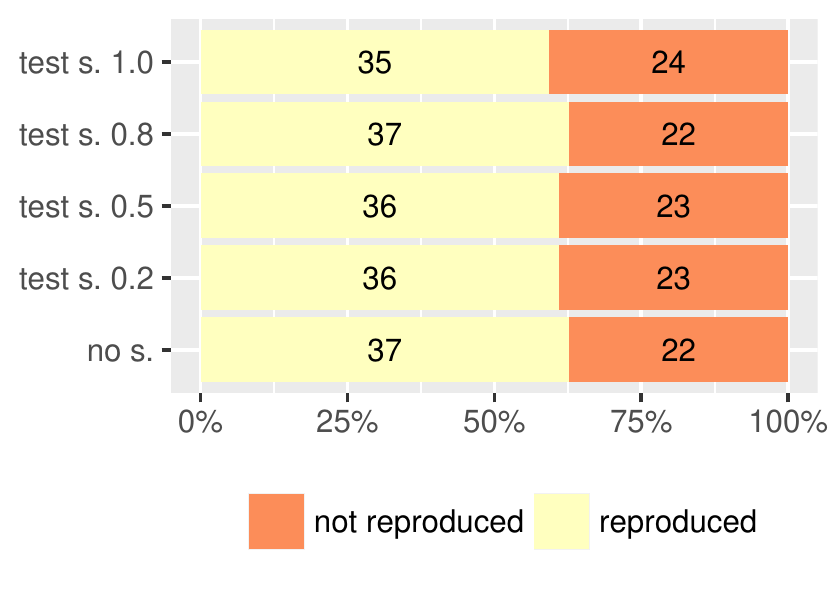}
	    \caption{test-seeding vs. no-seeding (for all projects together)}
	    \label{fig:eval:results-rq12}
	\end{subfigure}
	\begin{subfigure}[t]{0.48\textwidth}
	    \centering
	    \includegraphics[height=38mm]{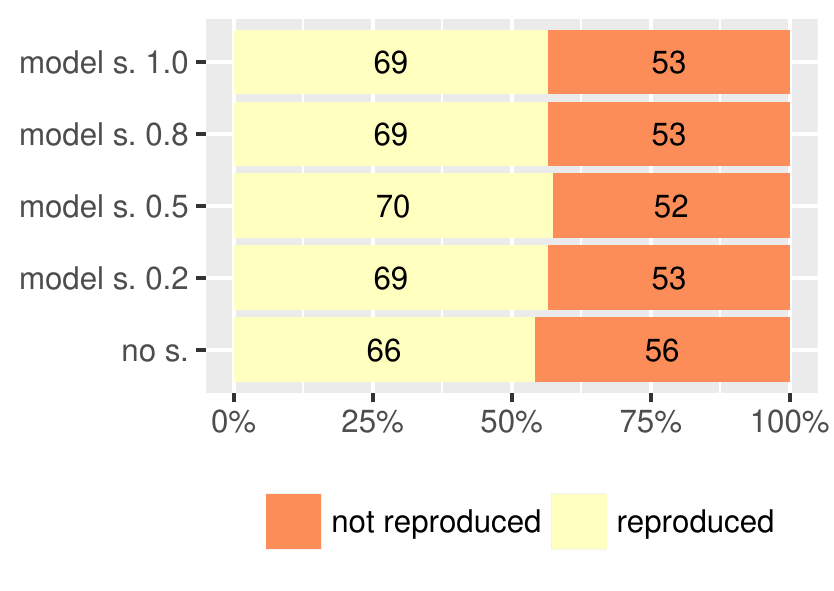}
	    \caption{model-seeding vs. no-seeding (for all projects together)}
	    \label{fig:eval:results-rq22}
	\end{subfigure}
%
%
	\begin{subfigure}[t]{0.48\textwidth}
	    \centering
	    \includegraphics[height=130mm]{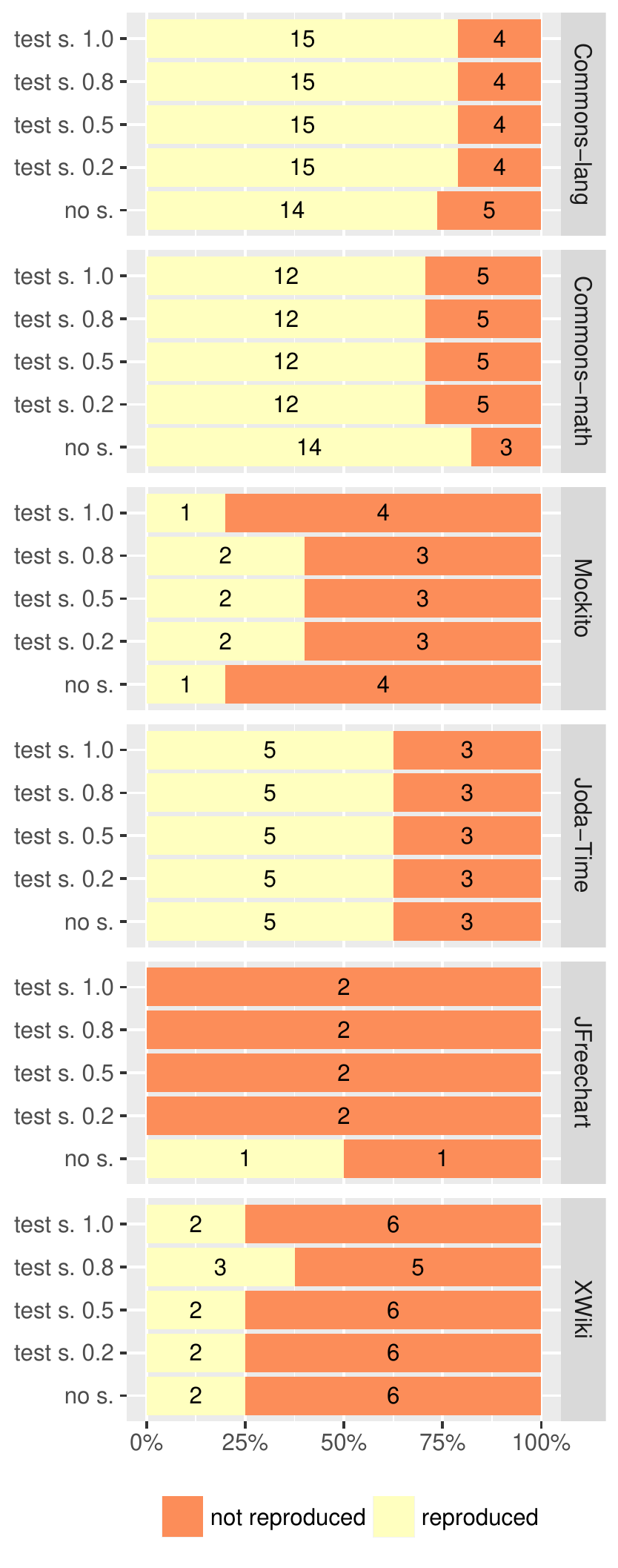}
	    \caption{test-seeding vs. no-seeding (per project)}
	    \label{fig:eval:results-rq12-apps}
	\end{subfigure}
	\begin{subfigure}[t]{0.48\textwidth}
	    \centering
	    \includegraphics[height=130mm]{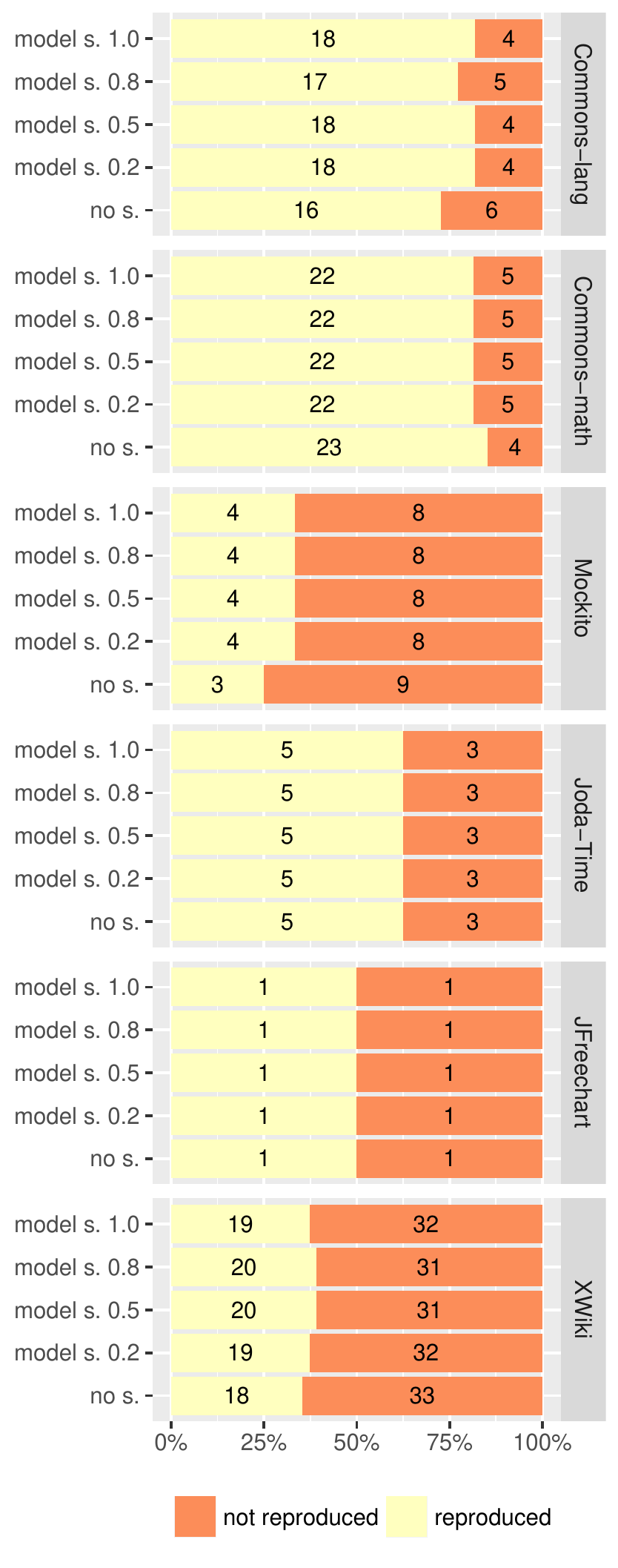}
	    \caption{model-seeding vs. no-seeding (per project)}
	    \label{fig:eval:results-rq22-apps}
	\end{subfigure}
	\caption{Outcomes observed in the majority of the executions for each crash in total and for each application.}
	\label{fig:eval:results}
\end{figure*}

Figure~\ref{fig:eval:results} demonstrates the comparison of each seeding strategy (left-side of the figure is for test seeding and right-side is for model seeding) with the baseline (no seeding). Figures~\ref{fig:eval:results-rq12} and \ref{fig:eval:results-rq22} show the overall comparison, while Figures~\ref{fig:eval:results-rq12-apps} and~\ref{fig:eval:results-rq22-apps} illustrate the per project comparison. In each of these figures, the yellow bar shows the number of reproduced crashes in the majority of the 30 executions, and the orange bar shows the non-reproduced crashes.

According to Figure \ref{fig:eval:results-rq12}, \textit{test s. 0.8} reproduced the same number of crashes. However, the other configurations of test-seeding reproduced fewer crashes in the majority of times. Moreover, according to Figure \ref{fig:eval:results-rq12-apps}, test seeding reproduces one more crash compared to no seeding. Also, some configurations of test seeding can reproduce one extra crash in  XWiki and commons-lang projects. On the contrary, all of the configurations of test seeding missed one and two crashes in JFreeChart and commons-math, respectively. Finally, we cannot see any difference between test seeding and no seeding in the Joda-Time project.


Table \ref{tab:crash-repr-table} demonstrates the impact of test-seeding on the crash reproduction ratio compared to no-seeding. It indicates that \textit{test s. 0.2 \& 0.5} have a better crash reproduction ratio for one of the crashes, while they perform significantly worse in 4 other crashes compared to no-seeding. The situation is almost the same for the other configurations of test seeding: \textit{test s. 0.8 \& 1.0} are significantly better in 2 crashes compared to no-seeding. However, they are significantly worse than no-seeding in 5 other crashes. The other interesting point in this table is the standard deviation crash reproduction ratio. This value is slightly higher for all of the test seeding configurations compared to no seeding. The values of odds ratios and and p-values for crashes with significant difference is available in Table \ref{tab:oddratios}.

The underlying reasons for the observed results in this section are analyzed in RQ1.4.

\begin{table} [t]
	\center
	\caption{Evaluation results for comparing seeding strategies (test and model seeding) and no-seeding in crash reproduction. $\overline{\text{ratio}}$ and $\sigma$  designate average crash reproduction ratio and standard deviation, respectively. The numbers in the comparison only count the statistically significant cases.}
	\label{tab:crash-repr-table}
    \begin{footnotesize}
	\begin{tabular}{ l r r | r r r }
\hline 
\textbf{Conf.} & \multicolumn{2}{c|}{Reproduction} & \multicolumn{3}{c}{Comparison to no s.} \\ 
  & $\overline{\text{ratio}}$ & $\sigma$ & better & worse \\ 
\hline 
test s. 1.0 & 23.7 & 11.01 & 2 & 5 \\ 
test s. 0.8 & 23.4 & 10.74 & 2 & 5 \\ 
test s. 0.5 & 23.8 & 10.76 & 1 & 4 \\ 
test s. 0.2 & 23.5 & 10.93 & 1 & 4 \\ 
no s. & 25.4 & 9.65 & - & - \\ 
\hline 
\end{tabular}
	\begin{tabular}{ l r r | r r r }
\hline 
\textbf{Conf.} & \multicolumn{2}{c|}{Reproduction} & \multicolumn{3}{c}{Comparison to no s.} \\ 
  & $\overline{\text{ratio}}$ & $\sigma$ & better & worse \\ 
\hline 
model s. 1.0 & 22.0 & 11.58 & 4 & 0 \\ 
model s. 0.8 & 21.9 & 11.92 & 4 & 1 \\ 
model s. 0.5 & 21.8 & 11.86 & 4 & 0 \\ 
model s. 0.2 & 21.6 & 12.00 & 3 & 1 \\ 
no s. & 21.3 & 12.32 & - & - \\ 
\hline 
\end{tabular}
    \end{footnotesize}
\end{table}

\subsubsection{Crash reproduction efficiency (\textbf{RQ1.2})}

Table \ref{tab:fitness-evaluation-table} demonstrates the comparison of test-seeding and no-seeding in the number of needed fitness function evaluations for crash reproduction. The average number of fitness function evaluations increases when using test-seeding. It means that test-seeding is slower than no-seeding on average. \textit{test s. 0.8} has the highest average fitness function evaluations. 

Moreover, the standard deviations of both no seeding and test seeding are high values (more than 20k evaluations). This notable variation is explainable due to the nature of search-based approaches. In some executions, the initialized population is closer to the objectives, and the search process can achieve reproduction faster. Similar variations are reported in the JCrashPack empirical evaluation as well~\cite{Soltani2019}. According to the reported standard deviations, we can see that this value increases for all of the configurations of test seeding compared to no seeding.

Also, the values of the effect sizes indicate that the number of crashes that receive (large or medium) positive impacts from \textit{test s. 0.2 \& 0.5} for their reproduction speed is higher than the number of crashes that exhibit a negative (large or medium) influence. However, this is not the case for the other two configurations. In the worst case, \textit{test s. 1.0} is considerably slower than no-seeding (with large effect size)  in 13 crashes.

\begin{table*} [t]
	\center
	\caption{Evaluation results for comparing test-seeding and no-seeding in the number of fitness evaluations $\overline{\text{evaluations}}$ and $\sigma$  designate average fitness function evaluations needed for crash reproduction and standard deviation, respectively. The numbers in the comparison only count the statistically significant cases.}
	\label{tab:fitness-evaluation-table}
	\begin{footnotesize}
	\begin{tabular}{ l r r | rr | rr | rr }
\hline 
\textbf{Conf.} & \multicolumn{2}{c|}{Fitness} & \multicolumn{6}{c}{Comparison to no s.} \\ 
  &   &   & \multicolumn{2}{c}{large} & \multicolumn{2}{c}{medium} & \multicolumn{2}{c}{small} \\ 
  & $\overline{\text{evaluations}}$ & $\sigma$ & $<0.5$ & $>0.5$ & $<0.5$ & $>0.5$ & $<0.5$ & $>0.5$ \\ 
\hline 
no s. & 10,467 & 22,368.13 & - & - & - & - & - & - \\ 
test s. 0.2 & 14,089 & 25,464& 4& 3& 1& 1& 2& - \\ 
test s. 0.5 & 13,366 & 25,043& 5& 3& 1& -& 2& 1 \\ 
test s. 0.8 & 14,254 & 25,496& 3& 4& 1& 5& 1& 3 \\ 
test s. 1.0 & 13,856 & 25,097& 3& 13& 4& 3& 1& 3 \\ 
\hline 
\end{tabular}
	\end{footnotesize}
\end{table*}

\subsubsection{Guided initialization effectiveness (\textbf{RQ1.3})}

Table \ref{tab:starting-effect-size} indicates the number of crashes where test-seeding had a significant (p-value $< 0.05$) impact on the search initialization compared to no-seeding. As we can see in this table, any configuration of test-seeding has a negative impact on the search starting process for 4 or 5 crashes. Additionally, this strategy does not have any significant beneficial impact on this aspect except on one crash in \textit{test s. 0.8}. Also, the standard deviation of the average search initialization ratios, in all of the configurations of test seeding, is increased compared to no seeding. For instance, this value for \textit{test s. 0.8} is about three times more than no seeding.

\begin{table}[t]
	\center
	\caption{Evaluation results for comparing seeding strategies (test and model seeding) and no-seeding in search initialization. $\overline{\text{ratio}}$ and $\sigma$  designate average successful search initialization ratio and standard deviation, respectively. The numbers in the comparison only count the statistically significant cases.}
	\label{tab:starting-effect-size}
	\begin{footnotesize}
	\begin{tabular}{ l r r | r r r }
\hline 
\textbf{Conf.} & \multicolumn{2}{c|}{Search started} & \multicolumn{3}{c}{Comparison to no s.} \\ 
  & $\overline{\text{ratio}}$ & $\sigma$ & better & worse \\ 
\hline 
test s. 1.0 & 26.9 & 9.22 & 0 & 5 \\ 
test s. 0.8 & 27.9 & 7.67 & 1 & 4 \\ 
test s. 0.5 & 26.9 & 9.22 & 0 & 5 \\ 
test s. 0.2 & 27.4 & 8.49 & 0 & 4 \\ 
no s. & 29.5 & 3.94 & - & - \\ 
\hline 
\end{tabular}
	\begin{tabular}{ l r r | r r r }
\hline 
\textbf{Conf.} & \multicolumn{2}{c|}{Search started} & \multicolumn{3}{c}{Comparison to no s.} \\ 
  & $\overline{\text{ratio}}$ & $\sigma$ & better & worse \\ 
\hline 
model s. 1.0 & 30.0 & 0.28 & 3 & 0 \\ 
model s. 0.8 & 30.0 & 0.00 & 3 & 0 \\
model s. 0.5 & 29.7 & 2.75 & 2 & 0 \\ 
model s. 0.2 & 29.5 & 3.87 & 2 & 1 \\
no s. & 29.2 & 4.72 & - & - \\ 
\hline 
\end{tabular}
	\end{footnotesize}
\end{table}

\subsubsection{Influencing factors (\textbf{RQ1.4})}
\label{sec:eval:rq14}

To finding the influencing factors in test seeding, we manually analyzed the cases which cause significant differences, in various aspects, between no-seeding and test-seeding. From our manual analysis, we identified 3 factors of the test seeding process that influence the search:
\begin{inparaenum}[(i)]
\item \textbf{Crash-Test Proximity},
\item \textbf{Crash-Object Proximity}, and
\item \textbf{Test Execution Cost}.
\end{inparaenum}

\paragraph{Crash-Test Proximity} For the first factor, we observe that \emph{cloning existing test cases} in the initial population leads to \emph{reproduce new crashes} when the cloned tests include elements which are close to the crash reproducing test. For instance, all of the configurations of test seeding are capable of reproducing the crash LANG 6b, while no-seeding cannot reproduce it. For reproducing this crash, Botsing needs to generate a string of a specific format, and this format is available in the existing test cases, which are seeded to the search process.

However, manually developed tests are not always helpful for crash reproduction. According to the results of Table \ref{tab:fitness-evaluation-table}, \textit{test s. 1.0}, which always clones test cases, is considerably and largely slower than no-seeding in 13 crashes. In these cases, cloning all of the test cases to form the initial population can misguide the search process to reach the crash reproducing test. As an example, Botsing needs to generate a simple test case, which calls the target method with an empty string and null object, to reproduce crash LANG-12b. But, \textit{test s. 1.0} clones tests which use the software under test in different ways.
To summarize, the overall quality of results of our test seeding solution is highly dependent on the quality of the existing test cases in terms of factors like the distance of existing test cases to the scenario(s) in which the crash occurs and the variety of input data.

\paragraph{Crash-Object Proximity} For the second factor, we observe that (despite the fixed value of $Pr[pick\ mut]$ for test seeding), the objects with call sequences carved from the existing tests and stored in the object pool can help during the search depending on their diversity and their distance from the call sequences that we need for reproducing the given crash. For instance, for crash MATH-4b, \botsing needs to initialize a \texttt{List} object with at least two elements before calling the target method in order to reproduce the crash. In test-seeding, such an object had been carved from the existing tests and allowed test seeding to reproduce the crash faster. Also, test-seeding can replicate this crash more frequently: the number of successfully replicated executions, in 30 runs, is higher with test-seeding.

In contrast, the carved objects can misguide the search process for some crashes which need another kind of call sequence.  For instance, in crash MOCKITO-9b, Botsing cannot inject the target method into the generated test because the carved objects do not have the proper state to instantiate the input parameters of the target method.

In summary, if the involved classes in a given crash are well-tested (the existing tests contain all of the usage scenarios of these classes), we have more chances to reproduce by utilizing test-seeding.

\paragraph{Test Execution Cost} The third factor points to the challenge of executing the existing test cases for seeding. The related tests for some crashes are either expensive (time/resource consuming) or challenging (due to the security issues) to execute. Hence, the \evosuite test executor, which is used by Botsing, cannot carve all of them. 

As an example of expensive execution, the EvoSuite test executor spends more than 1 hour during the execution of the related test cases for replicating frame 2 of crash Math-1b.

Also, as an example for security issues, the EvoSuite test executor is not successful in running some of the existing tests. It throws an exception during this task. For instance, this executor throws \texttt{java.lang.SecurityException} during the execution of the existing test cases for CHART-4b, and it cannot carve any object for seeding.

In some cases, test-seeding faces the mentioned problems during the execution of all of the existing test cases for a crash. If test seeding cannot carve any object from existing tests, there will be no useful call sequence in the object pool to seed during the search process. Hence, although the project contains some potentially valuable test scenarios for reproducing the given crash, there is no difference between no seeding and test seeding in these cases.


\subsubsection{Summary (\textbf{RQ1)}}

Test seeding (for any configuration) loses against no-seeding in the search initialization because some of the related test cases of crashes are expensive or even impossible to execute.
Also, we observe in the manual analysis that the lack of generality in the existing test cases prevents the crash reproduction search process initialization. In these cases, the carved objects from the existing tests mismatch the search process in the target method injection.
Moreover, this seeding strategy can outperform no seeding in the crash reproduction and search efficiency for some cases (\eg LANG 6b), thanks to the call sequences carved from the existing tests. However, these carved call sequences can be detrimental to the search process in some cases, if the carved call sequences do not contain beneficial knowledge about crash reproduction, overusing them can misguide the search process.

\subsection{Behavioral model seeding (RQ2)}

\subsubsection{Crash reproduction effectiveness (\textbf{RQ2.1})}

Figure~\ref{fig:eval:results-rq22} draws a comparison between model-seeding and no-seeding in the crash reproduction ratio according to the results of the evaluation on all of the 124 crashes. As mentioned in Section~\ref{sec:setup:setup:selection}, since model seeding collects call sequences both from source code and existing tests, it can be applied to all of the crashes (even the crashes that do not have any helpful test). As depicted in this Figure, all of the configurations of model-seeding reproduce more crashes compared to no-seeding in the majority of runs. We observe that \textit{model s. 0.2 \& 0.5 \& 1.0} reproduce 3 more crashes than no-seeding. In addition, in the best performance of model-seeding, \textit{model s. 0.8} reproduces 70 out of 124 crashes (6\% more than no-seeding).

Figure~\ref{fig:eval:results-rq22-apps} categorizes the results of Figure~\ref{fig:eval:results-rq22} per application. As we can see in this figure, model seeding replicates more crashes for XWiki, commons-lang, and Mockito. However, no-seeding reproduces one crash more than model-seeding for commons-math. For the other projects, the number of reproduced crashes does not change between no-seeding and different configurations of model-seeding. 

We also check how many crashes can be reproduced at least once with model seeding, but not with no seeding. In total, model-seeding configurations reproduce nine new crashes that no-seeding cannot reproduce.

Table \ref{tab:crash-repr-table} indicates the impact of model-seeding on the crash reproduction ratio. As we can see in this table, \textit{model s. 0.2} has a significantly better crash reproduction ratio in 3 crashes. Also, other configurations of model-seeding are significantly better than no seeding in 4 crashes. This improvement is achieved by model-seeding, while 2 out of 4 configurations of model-seeding have a significant unfavorable impact on only one crash. The values of odds ratios and and p-values for crashes with significant difference is available in Table \ref{tab:oddratios}.


\subsubsection{Crash reproduction efficiency (\textbf{RQ2.2})}

Table \ref{tab:fitness-evaluation-table-rq23} compares the  number of the needed fitness function evaluations for crash reproduction in model-seeding and no-seeding. As we can see in this table, the average effort is reduced by using model-seeding. On average \textit{mode s. 1.0} achieves the fastest crash reproduction.

According to this table, and in contrast to test-seeding, model-seeding's efficiency is slightly positive. The number of crashes that model-seeding has a positive large or medium influence (as Vargha Delaney measures are lower than 0.5) on varies between 3 to 5.
Also, model-seeding has a large adverse effect size (as Vargha Delaney measures are higher than 0.5) on one crash, while this number is higher for test-seeding (e.g., 13 for \textit{test s. 1.0}).

Table \ref{tab:fitness-evaluation-table-rq23} does not include the cost of model generation for seeding as mentioned in our experimental setup. In our case, model generation was not a burden and is performed only once per case study. We will cover this point in more detail in Section \ref{sec:discussion}.

\begin{table*} [t]
	\center
	\caption{Evaluation results for comparing model-seeding and no-seeding in the number of fitness evaluations $\overline{\text{evaluations}}$ and $\sigma$  designate average fitness function evaluations needed for crash reproduction and standard deviation, respectively. The numbers in the comparison only count the statistically significant cases.}
	\label{tab:fitness-evaluation-table-rq23}
	\begin{footnotesize}
	\begin{tabular}{ l r r | rr | rr | rr }
\hline 
\textbf{Conf.} & \multicolumn{2}{c|}{Fitness} & \multicolumn{6}{c}{Comparison to no s.} \\ 
  &   &   & \multicolumn{2}{c}{large} & \multicolumn{2}{c}{medium} & \multicolumn{2}{c}{small} \\ 
  & $\overline{\text{evaluations}}$ & $\sigma$ & $<0.5$ & $>0.5$ & $<0.5$ & $>0.5$ & $<0.5$ & $>0.5$ \\ 
\hline 
no s. & 18,713.1 & 28,023.93 & - & - & - & - & - & - \\ 
model s. 0.2 & 18,016.1 & 27,699.61& 2& 1& 1& 1& 2& 1 \\ 
model s. 0.5 & 17,646.9 & 27,463.02& 2& 1& 2& -& 2& 1 \\ 
model s. 0.8 & 17,564.5 & 27,400.27& 3& 1& 2& -& 1& 3 \\ 
model s. 1.0 & 17,268.8 & 27,190.73& 3& 1& 2& -& 1& 2 \\ 
\hline 
\end{tabular}
	\end{footnotesize}
\end{table*}

\subsubsection{Guided initialization effectiveness (\textbf{RQ2.3})}

Table \ref{tab:starting-effect-size}  provides a comparison between model-seeding and no-seeding in the search initialization ratio. As shown in this Table, \textit{model s. 0.2 \& 0.5} significantly outperform no seeding in starting the search process for two crashes. This number increases to 3 for \textit{model s. 0.8 \& 1.0}. In contrast to test-seeding, most of the configurations of model-seeding do not have any significant negative impact on the search initialization (only \textit{model s. 0.2} is significantly worse than no-seeding in one crash). Notably, the average search initialization ratios for all of the model seeding configurations are slightly higher than no seeding. In the best case for model seeding, \textit{model s. 0.8 \& 1.0} is 30/30 runs, and the standard deviations for these two configurations are 0 or close to 0.


\subsubsection{Influencing factors (\textbf{RQ2.4})}

We have manually analyzed the crashes which lead to significant differences between different configurations of model seeding and no seeding. In doing so, we have identified 4 influencing factors in model-seeding on search-based crash reproduction, namely:
\begin{inparaenum}[(i)]
\item using \textbf{Call sequence dissimilarity} for guided initialization,
\item having \textbf{Information source diversity} to infer the behavioral models,
\item \textbf{Sequence priority} for seeding by focusing on the classes involved in the stack trace, and
\item having \textbf{Fixed size abstract object behavior selection} from usage models.
\end{inparaenum}

\paragraph{Call sequence dissimilarity}
Using \emph{dissimilar call sequences} to populate the object pool in model seeding seems particularly useful for search efficiency compared to test seeding. In particular, if the number of test cases is large, model seeding enables to (re)capture the behavior of those tests in the model and regenerate a smaller set of call sequences which maximize diversity, augmenting the probability to have more diverse objects used during the initialization. For instance, Botsing with model-seeding is statistically more efficient than other strategies for replicating crash XWIKI-13141. Through our manual analysis we observed that model-seeding could replicate crash XWIKI-13141 in the initial population in 100\% of cases, while the other seeding strategies replicate it after a couple of iterations. In this case, despite the large size of the target class behavioral model  (35 transitions and 17 states), the diversity of the selected abstract object behaviors guarantees that Botsing seeds the reproducing test cases to the initial population.

\paragraph{Information source diversity}
Having \emph{multiple sources} to infer the model from helps to select diversified call sequences compared to test seeding. For instance, the sixth frame of the crash XWIKI-14556 points to a class called \texttt{HqlQueryExecutor}. No seeding cannot replicate this crash because it does not have any guidance from existing solutions. Also, since the test carver could not detect any existing test which is using the related classes, this seeding strategy does not have any knowledge to achieve reproduction. In contrast, the knowledge required for reproducing this crash is available in the source code, and model-seeding learned it from static analysis of this resource. Hence, this seeding strategy is successful in accomplishing crash reproduction.

\paragraph{Sequence priority}
By \emph{prioritizing classes} involved in the stack trace for the abstract object behaviors selection, the object pool contains more objects likely to help to reproduce the crash. For instance, for the 10th frame of the crash LANG-9b, model seeding could achieve reproduction in the majority of runs, compared to 0 for test and no seeding, by using the class \texttt{Fast\-Date\-Parser} appearing in the stack trace.

\paragraph{Fixed size abstract object behavior selection}
The last factor points to the fixed number of the generated abstract object behaviors from each model. In some cases, we observed that model-seeding was not successful in crash reproduction because the usage models of the related classes were large, and it was impossible to cover all of the paths with 100 abstract object behaviors. As such, this seeding strategy missed the useful dissimilar paths in the model. As an example, model-seeding was not successful in replicating crash XWIKI-8281 (which is replicated by no-seeding and test-seeding). In this crash, the unfavorable generated abstract object behaviors for the target class misguided the search process in model seeding.


\subsubsection{Summary (\textbf{RQ2})}

Model seeding achieves a better search initialization ratio compared to no seeding. With respect to the best achievement of model seeding (\textit{model s. 0.8 \& 1.0}), they decrease the number of not started searches in 3 crashes.
Moreover, compared to no seeding, model seeding increases the number of crashes that can be reproduced in the majority of times to 6\%. It also reproduces 9 (out of 124) extra crashes that are unreproducible with no-seeding. 
In addition, model seeding improves the efficiency of search-based crash reproduction compared to no seeding. It takes, on average, less fitness function evaluations. Also, model seeding delivers more positive significant impact on the efficiency of the search process compared to no seeding.

In general, model seeding outperforms no seeding in all of the aspects of search-based crash reproduction. According to the manual analysis that we have performed in this study, model seeding achieves this performance thanks to multiple factors: Call sequence dissimilarity, Information source diversity, and Sequence priority. Nevertheless, we observe a negative impacting factor in model seeding, as well. This factor is the fixed size abstract object behavior selection.


\section{Discussion}\label{sec:discussion}

\subsection{Practical implications}

\paragraph{Model derivation costs.} Generating seeds comes with a cost. 
For our worst case, XWIKI-13916, we collected 286K call sequences from static and dynamic analysis and generated 7,880 models from which we selected 6K abstract object behaviors. We repeated this process 10 times and found the average time for call sequence collection to be 14.2 seconds; model inference took 77.8 seconds; and abstract object behavior selection and concretization took 51.5 seconds. We do note however that the model inference is a one-time process that could be done offline (in a continuous integration environment). After the initial inference of models, any search process can utilize model seeding. To summarize, the total initial overhead is $\sim2.5$ minutes, and the total nominal overhead is around $\sim1.25$ minute.
We argue that \textbf{the overhead of model seeding is affordable giving its increased effectiveness. }
The initial model inference can also be incremental, to avoid complete regeneration for each update of the code, or limited to subparts of the application (like in our evaluation where we only applied static and dynamic analysis for classes involved in the stack trace).
Similarly, abstract object behavior selection and concretization may be prioritized to use only a subset of the classes and their related model. In our current work, this prioritization is based on the content of the stack traces. Other prioritization heuristics, based for instance on the size of the model (reflecting the complexity of the behavior), is part of our future work.



\paragraph{Applicability and effectiveness.} Generally, test seeding alone does not make crash reproduction more effective. Actually, test seeding has a more negative impact on the search-based crash reproduction.
Test seeding only uses dynamic analysis, which entails that it collects more accurate information from the potential usage scenarios of the software under test; it also means that this strategy collects more limited information for seeding. If these limited amounts of call sequences differ from the call sequences needed to reproduce the crash scenario, test seeding can misguide the crash reproduction search process.

In contrast to test-seeding, we observe that model seeding always performs better than no seeding with different configurations. As such, we observe that \textbf{model seeding can reproduce more crashes than other strategies}. Also, since model seeding also exploits test cases, thereby subsuming test seeding regarding the observed behavior of the application that is reused during the search, greater performance can be attributed to the analysis of the source code translated in the model.

In our experiments, various configurations of model seeding reproduced 8 new crashes that neither test seeding nor no seeding strategies could reproduce. Additionally, \textbf{only model seeding could reproduce stack traces with more than seven frames} (\eg LANG-9b). Still, model seeding missed the reproduction of one crash which is reproduced by no seeding.
Despite the achieved improvements by model seeding, this seeding strategy could not outperform no-seeding dramatically (crash reproduction improved by 6\%). To better understand the reasons for the results, we manually analyzed the logs of Botsing executions on the crashes for which model seeding could not show any improvements. Through this investigation, we noticed that the generated usage models in these cases are limited and they do not contain the beneficial call sequences for covering the particular path that we need for crash reproduction. The average size of the generated model in this study is 7 states and 14 transitions.
We believe that by collecting more call sequences from different sources (\ie log files), model seeding can increase the number of crash reproductions.

Also, we observe that the size of the generated abstract behaviors set is commensurate to the size of the inferred model. If we have a small model, and we choose too many abstract behaviors, we will get similar abstract behaviours that misguide the search process. In contrast, if we chose a small set of abstract behaviors from a behavioral model with a large size, we will miss the chance of using all of the potential of the model for increasing the chance of crash reproduction by the search process.

\paragraph{Extendability.} The usage models can be inferred from any resource providing call sequences. In this study, we used the call sequences derived from the source code and existing test cases. However, we can extend the models with extra resources (\eg execution logs). Also, the abstract object behavior selection approach can be adapted according to the problem. In this study, we used the dissimilarity strategy to increase the diversity of the generated tests.
Moreover, model seeding makes a distinction between using the object pool during guided initialization and guided mutation (as shown in Figure~\ref{fig:approach}). This distinction enables us to study the influence of seeding during the different steps of the algorithm independently. 


\subsection{Model seeding configuration}

\begin{table}[t]
	\center
	\caption{Evaluation results for comparing different configurations of model seeding in crash reproduction. $\overline{\text{rate}}$ and $\sigma$  designate average crash reproduction rate and standard deviation, respectively. The numbers in the comparison only count the statistically significant cases.}
	\label{tab:additional-expe-repr-table}
    \begin{footnotesize}
	\begin{tabular}{ l r r | r r r }
\hline 
\textbf{Conf.} & \multicolumn{2}{c|}{Reproduction} & \multicolumn{3}{c}{Comparison to other conf.} \\ 
  & $\overline{\text{rate}}$ & $\sigma$ & better & worse \\ 
\hline 
Pr[init]=0.0 Pr[mut]=0.3 & 18.8 & 13.81 & 0 & 11 \\ 
Pr[init]=0.0 Pr[mut]=0.6 & 19.0 & 13.64 & 0 & 10 \\ 
Pr[init]=0.0 Pr[mut]=0.9 & 19.0 & 13.55 & 0 & 13 \\ 
Pr[init]=0.2 Pr[mut]=0.0 & 20.4 & 12.42 & 2 & 2 \\ 
Pr[init]=0.2 Pr[mut]=0.3 & 19.6 & 12.87 & 0 & 7 \\ 
Pr[init]=0.2 Pr[mut]=0.6 & 19.8 & 12.88 & 1 & 5 \\ 
Pr[init]=0.2 Pr[mut]=0.9 & 19.4 & 13.15 & 0 & 7 \\ 
Pr[init]=0.5 Pr[mut]=0.0 & 20.8 & 12.17 & 3 & 1 \\ 
Pr[init]=0.5 Pr[mut]=0.3 & 20.6 & 12.29 & 3 & 2 \\ 
Pr[init]=0.5 Pr[mut]=0.6 & 19.4 & 13.24 & 0 & 7 \\ 
Pr[init]=0.5 Pr[mut]=0.9 & 20.0 & 12.58 & 1 & 5 \\ 
Pr[init]=0.8 Pr[mut]=0.0 & 21.8 & 11.46 & 8 & 0 \\ 
Pr[init]=0.8 Pr[mut]=0.3 & 21.6 & 11.53 & 6 & 0 \\ 
Pr[init]=0.8 Pr[mut]=0.6 & 21.8 & 11.77 & 8 & 0 \\ 
Pr[init]=0.8 Pr[mut]=0.9 & 20.8 & 11.96 & 3 & 2 \\ 
Pr[init]=1.0 Pr[mut]=0.0 & 21.6 & 11.53 & 6 & 0 \\ 
Pr[init]=1.0 Pr[mut]=0.3 & 23.0 & 11.31 & 12 & 0 \\ 
Pr[init]=1.0 Pr[mut]=0.6 & 21.6 & 11.82 & 8 & 0 \\ 
Pr[init]=1.0 Pr[mut]=0.9 & 22.6 & 11.30 & 11 & 0 \\ 
\hline 
\end{tabular}
    \end{footnotesize}
\end{table}

\begin{table*} [t]
	\center
	\caption{Evaluation results for comparing different configurations of model seeding in the number of fitness evaluations $\overline{\text{rate}}$ and $\sigma$  designate average fitness function evaluations needed for crash reproduction and standard deviation, respectively. The numbers in the comparison only count the statistically significant cases.}
	\label{tab:additional-expe-ff-evals-table}
	\begin{footnotesize}
	\begin{tabular}{ l r r | rr | rr | rr }
\hline 
\textbf{Conf.} & \multicolumn{2}{c|}{Fitness} & \multicolumn{6}{c}{Comparison to other configurations} \\ 
  &   &   & \multicolumn{2}{c}{large} & \multicolumn{2}{c}{medium} & \multicolumn{2}{c}{small} \\ 
  & $\overline{\text{evaluations}}$ & $\sigma$ & $<0.5$ & $>0.5$ & $<0.5$ & $>0.5$ & $<0.5$ & $>0.5$ \\ 
\hline 
Pr[init]=0.0 Pr[mut]=0.3 & 23,456.5 & 30,105.20& -& 5& -& 3& -& 3 \\ 
Pr[init]=0.0 Pr[mut]=0.6 & 23,066.3 & 29,976.23& -& 2& -& 5& -& 3 \\ 
Pr[init]=0.0 Pr[mut]=0.9 & 23,030.9 & 30,001.82& -& 7& -& 4& 1& 2 \\ 
Pr[init]=0.2 Pr[mut]=0.0 & 20,179.0 & 29,012.80& -& -& 1& 2& 1& - \\ 
Pr[init]=0.2 Pr[mut]=0.3 & 21,803.0 & 29,620.34& -& 2& 2& 5& -& 1 \\ 
Pr[init]=0.2 Pr[mut]=0.6 & 21,448.9 & 29,441.74& -& 2& -& 3& 1& 2 \\ 
Pr[init]=0.2 Pr[mut]=0.9 & 22,214.6 & 29,752.12& -& 2& 2& 5& 1& 1 \\ 
Pr[init]=0.5 Pr[mut]=0.0 & 19,371.3 & 28,668.58& -& -& 2& 1& 3& - \\ 
Pr[init]=0.5 Pr[mut]=0.3 & 19,766.8 & 28,849.00& -& -& 1& 2& 2& - \\ 
Pr[init]=0.5 Pr[mut]=0.6 & 22,245.2 & 29,729.80& -& -& -& 4& -& 3 \\ 
Pr[init]=0.5 Pr[mut]=0.9 & 21,030.0 & 29,302.03& -& 2& -& 3& 1& 2 \\ 
Pr[init]=0.8 Pr[mut]=0.0 & 17,329.0 & 27,693.98& 2& -& 6& -& -& 1 \\ 
Pr[init]=0.8 Pr[mut]=0.3 & 17,710.5 & 27,919.28& 1& -& 4& -& 3& - \\ 
Pr[init]=0.8 Pr[mut]=0.6 & 17,327.0 & 27,694.60& 2& -& 6& -& -& - \\ 
Pr[init]=0.8 Pr[mut]=0.9 & 19,383.3 & 28,659.38& -& -& 1& 1& 2& 2 \\ 
Pr[init]=1.0 Pr[mut]=0.0 & 17,730.5 & 27,906.92& 1& -& 4& -& 3& 1 \\ 
Pr[init]=1.0 Pr[mut]=0.3 & 14,863.9 & 26,275.53& 7& -& 3& -& -& - \\ 
Pr[init]=1.0 Pr[mut]=0.6 & 17,692.5 & 27,930.17& 2& -& 5& -& 1& - \\ 
Pr[init]=1.0 Pr[mut]=0.9 & 15,656.9 & 26,798.15& 7& -& 5& -& 1& - \\ 
\hline 
\end{tabular}
	\end{footnotesize}
\end{table*}

Model seeding can be configured with different $Pr[pick\ init]$ and $Pr[pick\ mut]$ probabilities. Like many other parameters in search-based test case generation~\cite{Arcuri2013}, the values of those parameters could influence our results.
Although a full investigation of the effect of $Pr[pick\ init]$ and $Pr[pick\ mut]$ on the search process is beyond the scope of this paper, we set up a small experiment on a subset of crashes (10 crash in total) with 15 new configurations, each one run 10 times.

Tables \ref{tab:additional-expe-repr-table} and \ref{tab:additional-expe-ff-evals-table} presents the configurations used for $Pr[pick\ init]$ and $Pr[pick\ mut]$ with, for each one, the crash reproduction effectiveness (Table \ref{tab:additional-expe-repr-table}), and the crash reproduction efficiency (Table \ref{tab:additional-expe-ff-evals-table}). In general, we observe that changing the probability of picking an object during guided initialization ($Pr[pick\ init]$) has an impact on the search and leads to more reproduced crashes with a lower number of fitness evaluations. This confirms the results presented in Section \ref{sec:eval-results}.
Changing the probability of picking an object during mutation ($Pr[pick\ mut]$) does not seem to have a large impact on the search.
A full investigation of the effects of $Pr[pick\ init]$ and $Pr[pick\ mut]$ on the search process is part of our future work.


\vspace{-0.5mm}

\subsection{Threats to validity}\label{sec:ttvalidity}

\subsubsection{Internal validity}
We selected 124 crashes from 5 open source projects: 33 crashes have previously been studied \cite{Soltani2018} and we added additional crashes from Xwiki and Defects4J (see Section~\ref{sec:setup}). 
Since we focused on the effect of seeding during guided initialization, we fixed the $Pr[pick\ mut]$ value (which, due to the current implementation of \botsing, is also used as $Pr[pick\ init]$ value in test seeding) to 0.3, the default value used in \evosuite for unit test generation. The effect of this value for crash reproduction, as well as the usage of test and model seeding in guided initialization, is part of our future work.
We cannot guarantee that our extension of \botsing is free of defects. We mitigated this threat by testing the extension and manually analyzing a sample of the results.
Finally, each frame has been run 30 times for each seeding configuration to take randomness into account and we derive our conclusions based on standard statistical tests \cite{Arcuri2014,Panichella2017c}.

\subsubsection{External validity}
We cannot guarantee that our results are generalizable to all crashes. However, we used JCrashPack, which is the most recent benchmark for Java crash reproduction. This benchmark is assembled carefully from seven Java projects and contains 200 real-life crashes. Since the \evosuite test executor is unsuccessful in running the existing test cases of one of the seven projects in JCrashPack (ElasticSearch), thereby test seeding and dynamic analysis of model-seeding are not applicable on crashes of this project, we excluded ElasticSearch crashes from JCrashPack. The diversity of crashes in this benchmark also suggests mitigation of this threat.

\subsubsection{Verifiability} 
A replication package of our empirical evaluation is available at \url{https://github.com/STAMP-project/ExRunner-bash/tree/master}. The complete results and analysis scripts are also provided in this package. Our extension of \botsing is released under a LGPL 3.0 license and available at \url{https://github.com/STAMP-project/botsing}.


\section{Future work}
\label{sec:future}

We observed that one of the advantageous factors in model seeding, which helps the search process to reproduce more crashes, consists in using more multiple resources for collecting the call sequences. Further diversification of sources is worth considering. In our future work, we will consider other sources of information, like logs of the running environment, to collect relevant call sequences and additional information about the actual usage of the application.

Also, collecting additional information from the log files would enable using full-fledged \emph{behavioral usage models} (\ie a transition system with probabilities on their transitions quantifying the actual usage of the application) to select and \emph{prioritize} abstract object behaviors according to that usage as it is suggested by statistical testing approaches \cite{Devroey2017b}.
For instance, we can put a high priority for the most uncommon observed call sequences for the abstract object behavior selection. We observed that selecting the most dissimilar paths in model-seeding helps the search process through crash reproduction. However, there is no guarantee that this approach is the best one. In future studies, we examine this approach with the new abstract object behavior selection approaches that we gain by the new full-fledged \emph{behavioral usage models}.

In this study, we focus on the impact of seeding during guided initialization by using different values for  $Pr[pick\ init]$ and $Pr[clone]$ and setting $Pr[pick\ mut]$ to the default value (0.3). However, our results show that even with the default value 0.3, using seeded objects during the search process helps to reproduce several crashes. Our future work includes a thorough assessment of that factor.
Furthermore, in the current version of model seeding, we noticed that the fixed size for the selected abstract object behaviors from the usage models could negatively impact the crash reproduction process. 
This set's size affects \botsing's performance and must be chosen carefully.  If too small, abstract object behaviors may not cover the transition system sufficiently, missing out on important usage information.  Too few abstract object behaviors can misguide the search process. In contrast, too many of them will lead to a time-consuming test concretization process.
In future investigations, we will study the integration of the search process with the abstract object behavior selection from the models. This integration can guide the seeding (\eg the abstract object behavior selection) using the current status of the search process.

Finally, we hypothesize that this seeding strategy may be useful for other search-based software testing applications and we will evaluate this hypothesis in our future work.


\section{Conclusion}
\label{sec:conclusion}

Manual crash reproduction is labor-intensive for developers.
A promising approach to alleviate them from this challenging activity is to automate crash reproduction using search-based techniques. In this paper, we evaluate the relevance of using both test and behavioral model seeding to improve crash reproduction achieved by such techniques. We implement both test seeding and the novel model seeding in \botsing.

For practitioners, the implication is that more crashes can be automatically reproduced, with a small cost. In particular, our results show that behavioral model seeding outperforms test seeding and no seeding without a major impact on efficiency. The different behavioral model seeding configurations reproduce 6\% more crashes compared to no seeding, while test seeding reduces the number of reproduced crashes. Also, behavioral model seeding can significantly increase the search initialization rate for 3 crashes compared to no seeding, while test seeding performs worse than no seeding in this aspect. We hypothesize that the achieved improvements by model seeding can be further extended by using more resources (\ie execution logs) for collecting the call sequences which are beneficial for the model generation.

From the research perspective, by abstracting behavior through models and taking advantage of the advances made by the model-based testing community, we can enhance search-based crash reproduction.
Our analysis reveals that (1) using collected call sequences, together with (2) the dissimilar selection, and (3) prioritization of abstract object behaviors, as well as (4) the combined information from source code and test execution, enable more search processes to get started, and ultimately more crashes to be reproduced.

In our future work, we will explore whether behavioral model seeding has further ranging implications for the broader area of search-based software testing. Furthermore, we aim to study the effect of changing the seeding probabilities on the search process, explore other sources of data to generate the model and try different abstract object behavior selection strategies.

\ack We would like to thank Annibale Panichella for his help and comments during the implementation of Botsing and the writing of this paper. This research was partially funded by the EU Horizon 2020 ICT-10-2016-RIA ``STAMP'' project (No.731529) and the Dutch 4TU project ``Big Software on the Run'' project.

\bibliographystyle{wileyj} 
\bibliography{biblio}

\end{document}